\documentclass[10pt]{article}

\usepackage{amsmath}
\usepackage{amssymb}
\usepackage{amsthm}
\usepackage{latexsym}
\usepackage{amscd}
\usepackage[mathscr]{eucal}

\newcommand{\dif}{\mathrm{d}} 
\newcommand{\eps}{\varepsilon}
\newcommand{\QED}{\raisebox{0.5ex}{\framebox[1.5ex]{\rule{0ex}{0ex}}}}
\newcommand{\R}{\mathbf{R}} 
\newcommand{\vol}{\mathrm{Vol}_{g}}
\newcommand{\mychi}{\raisebox{0.5ex}{$\chi$}}
\renewcommand{\div}{\mathrm{div}}
\newcommand{\Ric}{\mathit{Ric}}
\newcommand{\sym}{\mathcal{S}^{2}(\R^{3})}
\newcommand{\symtf}{\mathcal{S}_{g}^{2}(\R^{3})}
\newcommand{\jac}{\mathcal{J}(\R^{3})}

\newcommand{\oneform}{\Lambda^{1}(\R^{3})}
\newcommand{\Dif}{\mathrm{D}}
\newcommand{\bo}{\boldsymbol{0}}
\newcommand{\symtt}{\mathcal{S}_{TT}^{2} ( \R^{3})}
\newcommand{\symd}{\mathcal{S}_{\delta}^{2} ( \R^{3})}

\newtheorem*{mainthm}{Main Theorem} 
\newtheorem{thm}{Theorem}
\newtheorem{lemma}[thm]{Lemma} 
\newtheorem{cor}[thm]{Corollary}
\newtheorem{prop}[thm]{Proposition} 
\newtheorem{nonumthm}{Theorem}

\theoremstyle{definition}
\newtheorem{defn}[thm]{Definition}

\allowdisplaybreaks

\begin{document}

\title{Perturbative Solutions of the Extended Constraint Equations in
General Relativity}

\author{Adrian Butscher \\ 
Department of Mathematics \\ University of Toronto \\ E-mail: \ttfamily
butscher@utsc.utoronto.ca}

\maketitle

\renewcommand{\baselinestretch}{1}
\normalsize

\begin{abstract}
    The extended constraint equations arise as a special case of the
    conformal constraint equations that are satisfied by an initial
    data hypersurface $\mathcal{Z}$ in an asymptotically simple
    space-time satisfying the vacuum conformal Einstein equations
    developed by H. Friedrich.  The extended constraint equations
    consist of a quasi-linear system of partial differential equations
    for the induced metric, the second fundamental form and two other
    tensorial quantities defined on $\mathcal{Z}$, and are equivalent
    to the usual constraint equations that $\mathcal{Z}$ satisfies as
    a space-like hypersurface in a space-time satisfying Einstein's
    vacuum equation.  This article develops a method for finding
    perturbative, asymptotically flat solutions of the extended
    constraint equations in a neighbourhood of the flat solution on
    Euclidean space.  This method is fundamentally different from the
    `classical' method of Lichnerowicz and York that is used to solve
    the usual constraint equations.
\end{abstract}

\renewcommand{\baselinestretch}{1}
\normalsize

\section{Introduction}

The notion of an \emph{asymptotically simple} space-time was first
proposed by Penrose in the 1960s as a means for studying the
asymptotic properties of isolated solutions of Einstein's equations in
General Relativity \cite{penrose1}.  The central idea is to define a
class of space-times which are conformally diffeomorphic to the
interior of a Lorentz manifold with boundary, called the
\emph{unphysical space-time}, where the boundary is identified in a
certain way with points at infinity in the original space-time. 
Einstein's vacuum equations can be rephrased in terms of this
construction and can be used to describe the metric and conformal
boundary of the conformally rescaled space-time.  However, this
description is rather awkward in some respects because of the
following phenomenon.  The metric $g$ of the unphysical space-time is 
related to the metric $\tilde{g}$ of the original space-time by 
$\tilde{g} = \Omega^{-2} g$, where the \emph{conformal factor} 
$\Omega$ vanishes at the boundary (thereby encoding the fact that the 
boundary is at infinite distance).  Consequently, Einstein's equation 
in the unphysical space-time is $Ric(\Omega^{-2} g) = 0$, which 
degenerates at the boundary.

Helmut Friedrich has developed a new approach for the mathematical description of the unphysical space-time, known as the \emph{conformal Einstein equations}, which avoids this difficulty.  The reader is asked to consult \cite{jorg,fried3,fried4} for a review of these ideas.  Essentially, Friedrich has discovered equations which are equivalent to Einstein's equations in the unphysical space-time but which do not degenerate at the boundary.  Using these equations, Friedrich has made significant steps in the development of the Penrose model of asymptotic simplicity; in particular, he has been able to prove the first \emph{semi-global} existence and stability results for the Cauchy problem of evolving such space-times from initial data. Semi-global in this context means that, starting with so-called \emph{asymptotically hyperbolic} initial data, it is possible to generate the entire future development of the data under Cauchy evolution, all the way up to time-like infinity.  Recently, thanks to work of Corvino \cite{corvino} and Chru\'sciel and Delay \cite{piotr}, this result has been extended to a global existence result, in which a large class of asymptotically simple space-times can be constructed by Cauchy evolution from asymptotically flat initial data.

As is well known, the \emph{Cauchy problem} near a space-like hypersurface $\mathcal{Z}$ consists of splitting Einstein's equations into \emph{constraint equations} satisfied by initial data on $\mathcal{Z}$ and \emph{evolution equations} that describe how the initial data evolves in time to produce a space-time neighbourhood around $\mathcal{Z}$.  The conformal Einstein equations also admit such a Cauchy problem.  Indeed, given a space-like  hypersurface $\mathcal{Z}$ in the unphysical space-time, one can deduce a system of constraint equations and evolution equations for initial data on $\mathcal{Z}$.  The constraint equations on $\mathcal{Z}$ that arise in this way are called the \emph{conformal constraint equations}.  

The conformal constraint equations are far more complex than the `usual' constraint equations satisfied by the hypersurface $\mathcal{Z}$ when it is seen as an infinitely extended, space-like hypersurface of the original space-time.  The reason is that the conformal constraint equations combine two sub-structures --- namely `geometric' constraints (i.e.~resulting from the Gauss-Codazzi equations) and a boundary value problem at the boundary of $\mathcal{Z}$ --- in a highly coupled and non-linear way.  Although general techniques do not yet exist for analyzing these equations, there is one approach which may be fruitful: to tackle these two sub-structures individually and in isolation from one another by considering special cases.  The hope is that the knowledge gained in this way can be assembled in order to develop an understanding of the equations in more general cases.  Just such an approach has been commenced by the author in \cite{me3} and is extended in this article.

\paragraph*{The Extended Constraint Equations.} In the author's previous paper \cite{me3}, the \emph{Ansatz} $\Omega \equiv 1$ is used to eliminate the boundary value problem for $\Omega$ and bring the `geometric' constraint problem to the forefront.  This happens because the \emph{Ansatz} $\Omega \equiv 1$ essentially assumes that the conformal rescaling is trivial so that $\mathcal{Z}$ is a \emph{boundaryless} space-like hypersurface in the original space-time.  The system of equations which results from this \emph{Ansatz} reduces the conformal constraint equations considerably.  The only remaining unknowns are the induced metric $g$ and the second fundamental form $\chi$ of $\mathcal{Z}$, as well as the so-called \emph{electric} part $S$ and the \emph{magnetic} part $\bar S$ of the Weyl tensor $W$ of the space-time relative to the $3+1$ splitting induced by $\mathcal{Z}$.  (That is, $S_{ab} = W_{a00b}$ is the electric part of $W$ and $\bar{S}_{ab} = \eps^{st}_{\hspace{1.5ex} a} W_{b0st}$ is the magnetic part of $W$, where $\eps_{abc}$ is the fully antisymmetric permutation symbol associated to the metric $g$).  Note that both $S$ and $\bar S$ are trace-free and symmetric.   Explicitly, the conformal constraint equations reduce to the following system under the $\Omega \equiv 1$ \emph{Ansatz}, called the \emph{extended constraint equations}: 
\begin{equation}
    \label{eqn:ec}
    \begin{gathered}
	\nabla_{c} \chi_{ab} - \nabla_{b} \chi_{ac} = \eps^{e}_{\hspace{1ex}bc} \bar{S}_{ae}  \\
	\nabla^{a} \bar{S}_{ae} = \eps^{bc}_{\hspace{1.5ex} e} \chi^{a}_{b} R_{ac} \\
	R_{ab} = S_{ab} - \chi^{c}_{c} \chi_{ab} + \chi^{c}_{a} \chi_{cb} \\
	\nabla^{a} S_{ab} = - \chi^{ac} \eps^{e}_{\hspace{1ex} bc} \bar{S}_{ae}
    \end{gathered}
\end{equation}
where $\nabla$ is the covariant derivative of the metric $g$ and $R_{ab}(g)$ is its Ricci curvature.  (Actually, \eqref{eqn:ec} is a slightly different yet equivalent formulation of the extended constraint equations found in \cite{me3}.)

As equations involving the metric $g$ and the second fundamental form $\chi$ of the space-like hypersurface $\mathcal{Z}$ in the original space-time, one expects the extended constraint equations to be related in some way to the `usual' constraint equations satisfied by $g$ and $\chi$.  In fact, the following lemma shows that the extended constraint equations are fully equivalent to the usual constraint equations.

\begin{lemma}
	\label{lemma:same}
	Let $\mathcal{Z}$ be a space-like hypersurface in the original space-time having induced metric $g$ and second fundamental form $\chi$.  Then $g$ and $\chi$ satisfy the usual constraint equations on $\mathcal{Z}$ if and only if $g$ and $\chi$, along with two tensors $S$ and $\bar S$ defined via $S_{ab} \equiv R_{ab}(g) + \chi^c_c \chi_{ab} - \chi^c_a \chi_{cb}$ and $\bar S_{ae} \equiv \eps^{bc}_{\hspace{1.5ex} e} \nabla_c \chi_{ab}$ satisfy the extended constraint equations.
\end{lemma}
	
\begin{proof}
Suppose first that $(g, \chi, S, \bar{S})$ satisfy \eqref{eqn:ec}.  Then by taking traces in the first and third equations, one obtains the usual constraint equations, namely that $R = \chi^c_b \chi^b_c - (\chi_c^c)^2$ and $0 = \nabla^c \chi_{cb} - \nabla_b \chi^c_c$.

Next suppose that $g$ and $\chi$ satisfy the usual constraint equations.  Let  $S$ and $\bar{S}$ be defined as in the hypotheses of the lemma, in which case the desired symmetries of $S$ and $\bar S$ follow from the constraint equations and the first and third equations of \eqref{eqn:ec} follow automatically.  Some simple calculations are now needed to show that the second and fourth equations also hold.

First, the Bianchi identity for the Ricci curvature states $\nabla^{a} R_{ab} - \frac{1}{2} \nabla_{b} R = 0$.  Substitute for $R_{ab}$ and $R$ using the definition of $S$ to derive 
\begin{align}
    0 &= \nabla^{a} \big( S_{ab} - \mychi^{c}_{c} \mychi_{ab} +
    \mychi^{c}_{a} \mychi_{cb} \big) - \frac{1}{2} \nabla_{b} \big(- 
    (\mychi^{c}_{c})^{2} + \mychi^{ac} \mychi_{ac} \big) \notag \\
    &= \nabla^{a} S_{ab} - \big( \mychi^{c}_{c} \delta^{a}_{b} -
    \mychi^{a}_{b} \big) h^{uv} \bar{S}_{uav} + \mychi^{ac}
    \bar{S}_{abc}
    \label{eqn:vanish}
\end{align}
where $\bar S_{abc}$ is the dual of the tensor $\bar S_{ae}$ defined by $\bar S_{abc} \equiv \nabla_c \chi_{ab} - \nabla_b \chi_{ac}$.  By symmetry, the middle term in \eqref{eqn:vanish}
vanishes, leaving $0 = \nabla^{a} S_{ab} + \mychi^{ac} \bar{S}_{abc}$, which is exactly the fourth equation of \eqref{eqn:ec} when written in terms of this dual.

Next, consider the commutator of the second covariant derivatives of $\bar S$.  That is,
\begin{align}
    \eps^{ebc} \nabla_{e} \bar{S}_{abc} &= 2 \eps^{ebc} \nabla_{e}
    \nabla_{c} \mychi_{ab} \notag \\
    &= \eps^{ebc} \big( \nabla_{e} \nabla_{c} \mychi_{ab} - \nabla_{c}
    \nabla_{e} \mychi_{ab} \big) \notag \\
    &= \eps^{ebc} R_{eca}^{\hspace{3ex} s} \mychi_{sb}
    \label{eqn:subme}
\end{align}
using the symmetries of $R_{abcd}$.  Substitute in \eqref{eqn:subme} the well-known decomposition
$$R_{eca}^{\hspace{3ex} s} = g_{ea} R^{s}_{c} - \delta^{s}_{e} R_{ca}
+ \delta^{s}_{c} R_{ea} - g_{ca} R^{s}_{e} - \frac{1}{2} R \big(
g_{ea} \delta^{s}_{c} - \delta^{s}_{e} g_{ca} \big) \, ,$$
of the curvature tensor in three dimensions to obtain
\begin{equation}
    \eps^{ebc} \nabla_{e} \bar{S}_{abc} = 2 \eps^{bc}_{\hspace{2ex} a}
    \mychi^{s}_{b} R_{cs} \, .
    \label{eqn:equiv}
\end{equation}
The second equation of \eqref{eqn:ec} is now a consequence of \eqref{eqn:equiv} after invoking the symmetries of $\bar{S}$ once again.
\end{proof}

The proof of Lemma \ref{lemma:same} reveals an important feature of the extended constraint equations that will be exploited in a crucial way in the proof of the Main Theorem of this paper.  Indeed, the calculations above show that the extended constraint equations break up into two pairs of equations, each consisting of a primary `geometric' equation along with its \emph{integrability condition}, i.e.~an equation derived from the primary one by commuting covariant derivatives.  This is explicitly true of the second equation of \eqref{eqn:ec} but it is also true of the fourth equation since the Bianchi identity is the integrability condition for the Riemann curvature tensor.

\medskip \noindent \scshape Remark: \upshape Another way of thinking about this system of equations, given that $S$ and $\bar S$ are the electric and magnetic parts of the Weyl tensor, is that \eqref{eqn:ec} is equivalent to the vanishing of the full four-dimensional Ricci tensor along $\mathcal Z$.

\paragraph*{The Statement of the Main Theorem.} The extended constraint equations given in \eqref{eqn:ec} are still too complicated to solve for general initial data on general non-compact, boundaryless space-like hypersurfaces in an infinitely extended space-time.  An appropriate simplification is to consider \emph{perturbative} solutions near a given known solution.  An obvious known solution of the equations is the trivial solution in which $\mathcal{Z} = \R^3$ and $g = \delta$ (the Euclidean metric) along with $\chi = S = \bar{S} = 0$.  This paper will therefore investigate the nature of a certain class of perturbative solutions of the extended constraint equations decaying asymptotically to this trivial solution and `near' it in some appropriate sense.   In \cite{me3}, an approach for producing such solutions was developed under the further simplifying assumption of vanishing second fundamental form (the so-called \emph{time-symmetric} case).  The present paper will extend the approach used there to the full, non-time-symmetric case.  (Note: perturbations near general asymptotically flat metrics on more general asymptotically flat manifolds will not be considered here.)  The Main Theorem to be proved in this paper is the following.  

\begin{mainthm}
    There exists an infinite-dimensional Banach space $\mathcal{B}$ of so-called \emph{free data} consisting of triples $(T,\bar{T}, \phi)$, where $T$, $\bar{T}$ are symmetric, $\delta$-trace-free and $\delta$-divergence-free two-tensors on $\R^3$ and $\phi$ is a function on $\R^3$, which parametrizes solutions of the extended constraint equations near the trivial solution as follows.
\begin{enumerate}
	\item  For every triple $(T, \bar{T}, \phi)$ whose norm is sufficiently small, there is a unique solution $(g, \chi, S, \bar{S})$ of the extended constraint equations that is an asymptotically flat perturbation of the trivial solution $g = \delta$ and $\chi = S = \bar S = 0$ (in a sense to be defined later in this paper).
	
	\item The point $(0,0,0) \in \mathcal{B}$ corresponds to the trivial solution.  
	
	\item The association of triples $(T,\bar{T}, \phi)$ to solutions maps onto a neighbourhood of the trivial solution.
\end{enumerate}
Furthermore, the map taking $(T, \bar{T}, \phi)$ to the solution of the extended constraint equations is smooth in the sense of Banach spaces.
\end{mainthm}

\paragraph*{Sketch of Proof.} The Main Theorem will be proved by applying the Implicit Function Theorem to the extended constraint equations in a suitable way.  But because these equations are not \emph{fully-determined elliptic equations}, it is not possible to apply this theorem in a simple-minded way.  Rather, a two-step approach must be used.  The first step, called {\bf Theorem A} in this paper, consists of deriving and solving a \emph{related} system of equations which \emph{is} fully-determined elliptic.  This system will be called the \emph{secondary system}, and it will be shown that it determines the quantities $g$, $\chi$, $S$ and $\bar{S}$ uniquely in terms of certain free data in a Banach space $\mathcal{B}$ provided the norm of the free data is sufficiently small.  These solutions are of course not \emph{a priori} solutions of the extended constraint equations and thus it is still necessary to prove that a solution found in Theorem A is also a solution of the extended constraint equations.  This second step of the proof is called {\bf Theorem B} in this paper; it asserts that solutions of Theorem A \emph{are} solutions of the extended constraint equations. \hfill \qed

\medskip
A surprising technical difficulty arises in the proof of Theorem A.  In order to prove the existence of solutions via the Implicit Function Theorem, one must study the linearization of the secondary system at the trivial solution and show that it is bijective.  However, the fact of the matter is that this operator is injective but not surjective.  Consequently, it is possibly only to solve the secondary system up to an \emph{error term} belonging to the co-kernel of the linearization (which is finite-dimensional by Fredholm theory).  This error term may be non-zero, and its presence complicates the proof of Theorem B.  However, the lack of surjectivity is really an artifact of the two-step method of the proof of the Main Theorem, since the integrability conditions are exploited in a subtle way in Theorem B to prove that the potential error terms produced in Theorem A \emph{must} in fact be zero at the same time that the solution found in Theorem A does indeed satisfy the extended constraint equations.

The nature of the \emph{free data} of the Main Theorem can be characterized as follows.  Recall first that the  extended constraints are equivalent to the usual Einstein constraint equations, so that the method outlined above for solving the extended constraints can be interpreted as a completely new method, different from the `classical' Lichnerowicz-York method, for solving the usual constraint equations.  In the classical method (under the assumption of constant mean curvature), one freely prescribes a metric $g_{0}$ (without loss of generality whose components have unit determinant) and a symmetric tensor $\chi_{0}$, divergence free with respect to $g_{0}$, on $\R^{3}$.  The number of `freely prescribable' functions for data defined in this way is thus seven, or four if one first makes a specific choice of coordinates.  Then, one considers the conformally rescaled metric $g = u^{4} g_{0}$, where $u : \R^{3} \rightarrow \R$ is an unknown function, and reduces the constraint equations to a semi-linear elliptic equation for $u$.  Once this equation is solved, a further rescaling of $\chi_{0}$ yields a tensor $\chi$ which is trace-free and divergence-free with respect to $g$.   The Lichnerowicz-York method thus produces a solution $(g, \chi)$ in terms of the four degrees of freedom represented by $g_0$ and $\chi_0$. In contrast, the present method treats the metric $g$ and certain components of $\chi$, $S$ and $\bar{S}$ as the unknowns and leads to a quasi-linear elliptic system for these quantities that will be solved in terms of totally different free data.  In the derivation of the secondary system in Section 2.4, it will be shown that the free data consists of certain components of the curvature of $g$, amounting to four degrees of freedom, along with a single degree of freedom corresponding to the trace of the second fundamental form (which is absent in the Lichnerowicz-York solution since the mean curvature is kept constant there). 

\paragraph*{Acknowledgements.} The research and most of the writing for this paper was carried out while I was a post-doctoral fellow at the Max Planck Institute for Gravitational Physics in Golm, Germany.  I would like to thank Helmut Friedrich of this institute for suggesting the topic of the extended constraint equations as well as for suggesting the two-step approach for its solution.  I would also like to thank him for providing guidance and support during and after the course of the research.  Furthermore, I would like to thank Piotr Chru\'sciel, Justin Corvino, J\"org Frauendiener and Gerhard Huisken for their helpful suggestions and comments. 

\paragraph*{Erratum.} Note that \cite{me3} contains an error in the section pertaining to that paper's analogue of Theorem B.  The Stability Lemma in section 10.3.6 of \cite{me3} is mis-stated and its proof is incorrect.   It is then applied to the associated equations in the way the correctly-stated version should have been applied.  Since the results of this paper subsume and generalize those of \cite{me3}, the Main Theorem of \cite{me3} is still valid.

\section{Preliminaries}

\subsection{Spaces of Tensors and Tensor Operators}

For the sake of convenience, it is worthwhile to assign names to the various tensors spaces and tensor operators that appear in \eqref{eqn:ec} and elsewhere in this paper.  

Suppose $M$ is any Riemannian manifold with metric $g$.  Let $\mathcal{S}(M)$ denote the space of symmetric two-tensors over $M$ and let $\mathcal{S}_g(M)$ denote the symmetric two-tensors of $M$ that are in addition trace-free with respect to $g$.  Additionally, let $\Lambda^1(M)$ denote the one-forms of $M$.  Finally, let  $\mathcal{J}(M)$ denote those three-tensors which possess the symmetries $J_{abc} + J_{cab} + J_{bca} = 0$ and $J_{abc} + J_{acb} = 0$.  Such tensors will make an appearance at one point in this paper.  Next, define the following operators:
\begin{itemize}
   
    \item $\div_{g} :  \mathcal{S}^2 (M) \rightarrow \Lambda^1(M)$, given in local coordinates
    by $[\div_{g} (S)]_{b} = \nabla^{a} S_{ab}$;
        
    \item $\Ric :  \mathcal{S}^2 (M) \rightarrow \mathcal{S}^2 (M)$, given by in local coordinates by $[\Ric(g)]_{ab} =
    R_{ab}(g)$.  This is naturally only defined on the \emph{non-degenerate}
    symmetric tensors.
\end{itemize}
These operators appear in the system \eqref{eqn:ec}.  The following additional operators will play an important role in the sequel.  First, there are the operators
\begin{itemize}

    \item $\mathcal{D}_{g} : \mathcal{S}^2 (M) \rightarrow \mathcal{J}(M)$, given in local coordinates by
    $[\mathcal{D}_{g} (\chi)]_{abc} = \nabla_{c} \chi_{ab} -
    \nabla_{b} \chi_{ac}$;

    \item $\mathcal{Q}_{g} : \mathcal{J}(M) \rightarrow \Lambda^1(M)$ defined in local coordinates
    by $[\mathcal{Q}_{g} (J)]_{a} = \eps^{ebc} \nabla_{e} J_{abc}$,
\end{itemize}
which will be used in exploiting the integrability conditions built into the extended constraint equations.  Second, there are the formal adjoints (up to numerical factors) of the operators
$\mathcal{D}_{g}$ and $\div_{g}$, given by:
\begin{itemize}
    \item $\mathcal{D}_{g}^{\ast} : \mathcal{J}(M) \rightarrow \mathcal{S}_g(M) $, given by
    $[\mathcal{D}_{g}^{\ast} (J) ]_{ac}= \nabla^{b} J_{abc} + \nabla
    ^{b} J_{cba} - \frac{2}{3} \nabla^{b} J_{ubv} g^{uv} g_{ac}$. 
    Note that the adjoint of the restriction of the operator
    $\mathcal{D}_{g}$ to the space $\mathcal{S}_g(M) $ is being given here.
    
    \item $\div_{g}^{\ast} : \Lambda^1(M) \rightarrow \mathcal{S}_g(M) $, given by
    $[\div_{g}^{\ast}(X)]_{ac} = \nabla_{a} X_{c} + \nabla_{c} X_{a} -
    \frac{2}{3} \nabla^{b} X_{b} g_{ac}$.  Note that this operator is
    also known as the \emph{conformal Killing operator}.
\end{itemize}
It is a simple matter of calculating these adjoints by means of the
integration by parts formula for covariant derivatives and so this will not
be carried out here.

\subsection{Weighted Sobolev Spaces}

In order to proceed with the solution of the extended constraint equations, it is first necessary to specify in what Banach spaces the various unknown quantities lie.  The notion of asymptotic flatness in $\R^{3}$ should be encoded rigorously into these spaces by requiring that the relevant objects belong to a space of tensors with built-in control at infinity.  Furthermore, the spaces should be chosen to exploit the Fredholm properties of the operators above.  Both these ends will be served by weighted Sobolev spaces, whose definition and some of whose properties will be given in this section as a reminder to the reader (details can be found in such works as \cite{bartnik, cantor,cbc, com,lock1,mcowen}).  The actual choice of Banach spaces in which the solutions of the Main Theorem will be found will then be given in the next section.

Let $T$ be any tensor on $\R^{3}$.  (This tensor may be of any order --- the norm $\Vert \cdot \Vert$ appearing in the following definition is then simply the norm on such tensors that is induced from the
Euclidean metric of $\R^{3}$).  Note that the following definitions can also be made for general metrics on $\R^{3}$ but this will not be necessary in this paper.)  The $H^{k,\beta}$ Sobolev norm of $T$ is the quantity
$$\Vert T \Vert_{H^{k,\beta}} = \left( \sum_{l=0}^{k} \int_{\R^{3}}
\Vert D^{l} T \Vert^{2} \sigma^{-2(\beta-l) -3} \right)^{1/2} \, ,$$
where $\sigma(x) = ( 1+ r^{2} )^{1/2}$ is the \emph{weight function} and $r^{2} = (x^{1})^{2} + (x^{2})^{2} + (x^{3})^{2}$ is the squared distance to the origin.  Note that Bartnik's convention \cite{bartnik} for the power of $\sigma$ in the definition of the weighted spaces is being used (the reason for this is psychological: if $f \in H^{k,\beta}$ and $f$ is smooth enough to invoke the Sobolev Embedding Theorem (again, see \cite{bartnik}), then $f(x) = o(r^{\beta})$ as $r \rightarrow \infty$, which is easy to remember).

The space of $H^{k,\beta}$ functions of $\R^{3}$ will be denoted by $H^{k,\beta}(\R^{3})$ and the space of $H^{k,\beta}$ sections of a tensor bundle $B$ over $\R^{3}$ will be denoted by $H^{k,\beta} (B)$.  As an abbreviation, or where the context makes the bundle clear, such
a space may be indicated simply by $H^{k,\beta}$.  Note also that the following convention for integration will be used in the rest of this paper.  An integral of the form $\int_{\R^{3}} f$, as in the
definition above, denotes an integral of $f$ with respect to the standard Euclidean volume form.  Integrals of quantities with respect to the volume form of a different metric will be indicated explicitly,
as, for example, $\int_{\R^{3}} f \, \dif \vol$.

Constant-coefficient elliptic partial differential operators acting on spaces of $H^{k,\beta}$ tensors satisfy several important analytic properties, and two of these will be used in a crucial way in the sequel.  The first property is a characterization of the kernels and co-kernels of such operators that will be used in Theorem A.  The second property concerns the stability of the co-kernels of these operators that will be used in Theorem B.

\begin{prop}[Kernel/Co-Kernel] 
\label{prop:inv}
Let $B$ be a tensor bundle over $\R^{3}$ and $Q: H^{k,\beta}(B) \rightarrow H^{k-2, \beta - 2}(B)$ any linear, second order, homogeneous, elliptic partial differential operator with constant coefficients, where $\beta \not\in \mathbf{Z}$, and $k\geq 2$.  Denote by $\mathit{Ker}(Q : \beta)$ the kernel of $Q$ acting on $H^{k,\beta}$.    Then the following is true:
\begin{enumerate}
    	\item $Q$ is Fredholm;
	
	\item $Q$ is injective if $\beta < 0$ and surjective if $\beta > -1$;  
	
	\item If $\beta > 0$, then $\mathit{Ker}(Q : \beta)$ consists of polynomials of degree equal to the integer part of $\beta$ and has non-zero, finite dimension.
	
	\item If $\beta < -1$, then the image of $Q$ is the space
	    $$\mathrm{Im}(Q) = \left\{ y \in H^{k-2, \beta - 2}(B) \, : \,
	    \int_{\R^{3}} \langle y,z \rangle = 0 \quad \forall \, z \in Ker
	    (Q^{\ast} : -\beta -1) \right\} \, ,$$
	    where $Q^{\ast}$ is the formal adjoint of $Q$.  The image thus has non-zero finite codimension
\end{enumerate}
\end{prop}

\begin{proof} 
    The fact that the operator $Q$ is Fredholm and has finite dimensional kernel and co-kernel is a classical result that can be found in \cite{lock1,mcowen2}.  Standard Schauder theory then asserts that any solution of the equation $Q(u) = 0$ is smooth, and if it has polynomial growth or decay at infinity, then it is a polynomial or is zero, respectively.  The characterization of the image of $Q$ as the orthogonal complement of the kernel of the adjoint $Q^\ast$ is elementary functional analysis.  An excellent source for understanding the motivation behind this theorem can be found in \cite{mcowen} in which the behaviour of the Laplace operator on $\R^n$ is explained.
\end{proof}

\begin{prop}[Stability] 
    \label{prop:stab}
    Suppose $B$ is any tensor bundle over $\R^3$   
    and let $Q_{\eps}:H^{k,\beta} (B) \rightarrow H^{k-2,\beta-2}
    (B)$, $\eps \in [0,1]$, be a continuous family of linear,
    elliptic operators.  Furthermore, suppose that $Q_{\eps}$ is
    uniformly injective for any $\eps$; i.~e.\ there is a constant $C$
    independent of $\eps$ so that $\Vert Q_{\eps}(y)
    \Vert_{H^{k-2,\beta-2}} \geq C \Vert y \Vert_{H^{k, \beta}}$. 
    Finally, suppose $\mathcal{C}$ is a finite-dimensional linear
    subspace of $H^{k-2, \beta-2}(B)$.  If $\mathcal{C} \cap
    \mathrm{Im}(Q_{0}) = \{0\}$, then there exists $\eps_{0} > 0$ so
    that $\mathcal{C} \cap \mathrm{Im }(Q_{\eps}) = \{0\}$ for all
    $\eps < \eps_{0}$.
\end{prop}
    
\begin{proof} 
    
Suppose that this proposition is false; that is, for every $\eps > 0$, let $z_{\eps}$ be a non-zero element of $\mathcal{C} \cap \mathrm{Im} (Q_{\eps})$ and without loss of generality, it is possible to take $\Vert z_{\eps} \Vert_{H^{k-2, \beta-2}} =1$.  Since $z_{\eps} \in \mathcal{C}$ for every $\eps$, the finite-dimensionality of $\mathcal{C}$ implies that there is a subsequence $\eps_{j} \rightarrow 0$ and $z_{\eps_{j}} \equiv z_{j}$ that converges in the $H^{k-2, \beta -2}$ norm to a non-zero element $z \in \mathcal{C}$, with $\Vert z \Vert_{H^{k-2, \beta -2}} = 1$.  Furthermore, since $z_{j} \in \mathrm{Im}(Q_{j})$, there is an element $y_{j} \in H^{k,\beta}(B)$ so that $Q_{j} (y_{j}) = z_{j}$.

\medskip \noindent \scshape Claim: \upshape there is $y \in H^{k,\beta}(B)$ so that $y_{j} \rightarrow y$ and $Q_{0}(y) = z$.

\medskip \noindent First, by the uniform injectivity of $Q_{j}$,
\begin{equation}
    \Vert y_{j} \Vert_{H^{k,\beta }} \leq C \Vert z_{j} 
    \Vert_{H^{k-2, \beta -2}} = C \, ,
    \label{eqn:unifybd}
\end{equation}
so that the sequence $y_{j}$ is uniformly bounded in $H^{k,\beta}$.  Next, by the injectivity of the operator $Q_{0}$,
\begin{align}
    \label{eqn:cauchy}
    \Vert y_{i} - y_{j} \Vert_{H^{k, \beta}} &\leq C \Vert Q_{0} 
    (y_{i} - y_{j} ) \Vert_{H^{k-2, \beta-2}} \notag \\
    &\leq C \big( \Vert (Q_{0} - Q_{i} ) ( y_{i}) \Vert_{H^{k-2,
    \beta-2}} + \Vert Q_{i} (y_{i}) - Q_{j} (y_{j}) \Vert_{H^{k-2,
    \beta-2}}  \notag \\
    &\qquad + \Vert (Q_{0} - Q_{j} ) ( y_{j}) \Vert_{H^{k-2, \beta-2}}
    \big) \notag \\
    &\leq C \big( \Vert Q_{0} - Q_{i} \Vert_{op} + \Vert z_{i} - z_{j}
    \Vert_{H^{k-2, \beta-2}} + \Vert Q_{j} - Q_{0} \Vert_{op} \big)
\end{align}
by \eqref{eqn:unifybd}, where $\Vert \cdot \Vert_{op}$ denotes the operator norm in the space of linear operators on $H^{k-2, \beta-2} (B)$.  Now the first and last terms in \eqref{eqn:cauchy} go to zero with sufficiently large $i$ and $j$ because the family $Q_{\eps}$ is continuous, while the middle term goes to zero by construction.  Hence the sequence $y_{j}$ is Cauchy in $H^{k-2, \beta-2}$ and thus converges to an element $y \in H^{k-2, \beta-2}(B)$.  By similar estimates as above, it is straightforward to show that $\Vert z - Q_{0} (y) \Vert_{H^{k-2, \beta-2}}$ is zero.  Thus $z = Q_{0}(y)$.   But $z$ is a non-zero element in $\mathcal{C} \cap \mathrm{Im}(Q_{0})$.  This contradicts the hypotheses of the theorem. 
\end{proof}

\subsection{Choosing the Banach Spaces}

Solutions of the extended constraint equations will be found in the
following Banach spaces.  Pick any $\beta \in (-1,0)$ and any $k \geq
4$.  Then choose:
\begin{itemize}
    \item metrics $\delta + h$ so that $h \in H^{k,\beta} \big( \sym
    \big)$;
    
    \item tensors $\chi$ in $H^{k-1, \beta-1} \big( \sym \big)$;
    
    \item tensors $S$ and $\bar{S}$ in $H^{k-2, \beta-2} \big( \symtf
    \big)$.
\end{itemize}

The preceding choice of Banach spaces is necessitated by the following two considerations.  First, in order to ensure that the metric $\delta + h$ is asymptotically flat, $h$ must decay as $r \rightarrow \infty$, and this holds by the Sobolev Embedding Theorem when $\beta <0$.  Next, a non-trivial, asymptotically flat metric satisfying the constraint equations must satisfy the Positive Mass Theorem \cite{sy} and consequently must have non-zero ADM mass.  Thus the $r^{-1}$ term in the asymptotic expansion of $h$ must be allowed to be non-zero, which imposes the further requirement that $\beta > -1$.  Then, as a consequence of the choice made for $h$, the $\chi$, $S$ and $\bar{S}$ quantities must be chosen as above because of the differing numbers of derivatives taken on these quantities in the equations \eqref{eqn:ec}: since the equations are meant to define maps between weighted Sobolev spaces, the weightings on $\chi$ and $S$ and $\bar{S}$ must match together properly and match the weighting on the metric $\delta + h$.  (For example, the Ricci curvature operator is of degree two and sends a metric $\delta + h$ with $h \in H^{k,\beta}$ to a tensor in $H^{k-2, \beta-2}$.  Thus $S$
must lie in $H^{k-2, \beta -2}$ to match $\Ric(\delta + h)$.)  Finally,  $k \geq 4$ is required in order to apply the Sobolev Embedding Theorem at one stage of the proof of the Main Theorem.

\medskip \noindent \scshape Remark: \upshape One issue has been glossed over in the previous paragraph, and this is the effect of the non-linear terms.  Because of this, for example, it is not immediately obvious that $\Ric(\delta + h)$ is in $H^{k-2, \beta -2}$ when $h \in H^{k,\beta}$ because this expression involves products of the metric and its first and second derivatives.  However, the Multiplication Theorem for weighted Sobolev spaces \cite{cbc} implies that the choice for $\beta$ and $k$ made above ensures that the various Banach spaces are in fact Banach algebras, and so this is not a problem.

\subsection{The Secondary System To Be Used in Theorem A}

The first step of the proof of the Main Theorem is to derive and solve a \emph{secondary system}  of equations that is  related to the extended constraint equations in a clever way.  The key modification is that the secondary system is fully-determined and elliptic.   The secondary system will be given immediately to streamline the presentation, whereas the details of its motivation and derivation will be given thereafter.   

The objects that will appear in the definition of the secondary system are these. 
\begin{enumerate}
	\item Let $\mathcal{S}_{TT} (\R^{3})$ denote the space of symmetric, $\delta$-trace-free and $\delta$-divergence-free two-tensors and define the map 
\begin{equation}
	\label{eqn:yorkdecomp}
	\mathcal{Y} : H^{k-1,\beta -1} \big( \oneform \big) \times H^{k-2,\beta-2} \big(\symtt \big) \rightarrow H^{k-2, \beta -2} \big( \symtf \big)
\end{equation}
by $\mathcal{Y} (V, W) = \div^\ast_g (V) + W - \frac{1}{3}  \mathit{Tr}_g (W) g$.  Then put $S(X,T) = \mathcal{Y}(X, T)$ and $\bar{S} (\bar{X}, \bar{T}) = \mathcal{Y}(\bar X, \bar T)$.

	\item Define the map $\chi : H^{k-1, \beta - 1} \big( \symd \big) \times
H^{k-1, \beta -1} (\R^{3}) \rightarrow H^{k-1, \beta - 1} \big( \sym \big)$ via $\chi(K, \phi) = K + \tfrac{1}{3} \phi \delta$.

	\item Let $\Ric^H(g)$ denote the formal expression of the Ricci curvature of a metric $g$ in harmonic coordinates.  In local coordinates, this is given by 
\begin{equation}
	\label{eqn:redric}
	R_{ab} = R^{H}_{ab} + \frac{1}{2} (\Gamma_{a;b} + \Gamma_{b;a}) \, ,
\end{equation}
where $\Gamma^a = g^{st} \Gamma_{st}^a$ and $\Gamma_{st}^a$ are the Christoffel symbols of $g$.  
\end{enumerate}
The secondary system can now be defined as follows.

\begin{defn} 
	\label{defn:sec} 
	Let $g = \delta + h$ be the metric of $\R^3$ and let $S = S(X, T)$, $\bar{S} = \bar{S}(\bar{X}, \bar{T})$ and $\chi = \chi(K, \phi)$ be as above.  Then the \emph{secondary system} is:  
	\begin{equation}
	    \label{eqn:assoc}
	    \begin{gathered}
		\left[ \mathcal{D}_{g}^{\ast} \big( \mathcal{D}_{g} (\chi) -
		\star \, \bar{S} \big) \right]_{ab} = 0\\
		\left[\div_{g} (\bar{S}) \right]_{a} -
		\eps^{st}_{\hspace{1.5ex} a} \chi^{u}_{s} R_{ut}(g) = 0 \\
		\left[ \Ric^{H}(\delta + h) \right]_{ab} - S_{ab} +
		\chi^{s}_{s} \chi_{ab} - \chi^{s}_{a} \chi_{bs} = 0 \\
		\left[ \div_{g} (S) \right]_{a} + \chi^{st} \eps^{u}_{
		\hspace{1.5ex} at} \bar{S}_{su} = 0
	    \end{gathered}
	\end{equation}
	Where $\star \bar{S}$ denotes the tensor given in local coordinates by $\bar{S}_{ae} \eps_{\hspace{1ex} bc}^{e}$. 
\end{defn}

\paragraph*{Derivation of the Secondary System.} Each equation of \eqref{eqn:ec} has a corresponding equation in \eqref{eqn:assoc} and the motivation behind each of these equations  will be given in turn.

Begin by considering the first equation of \eqref{eqn:ec}.  In \cite{me3}, it was found that the symbol of this equation has a one dimensional kernel consisting of symmetric, pure-trace tensors.  The symbol is thus injective when restricted to the space of symmetric, trace-free tensors and thus the equation is over-determined elliptic when so restricted.  A related fully-determined elliptic equation can thus be derived as follows.  First, one decomposes $\chi$ into its trace and trace-free parts (with respect to the metric $\delta$) as $\chi = K + \frac{1}{3} \phi g$ and substitutes this into the first equation of \eqref{eqn:ec}.  One then composes both sides of this equation with the adjoint of the operator $\mathcal{D}_{g}$, so that the resulting equation contains the operator $\mathcal{D}_g^\ast \circ \mathcal{D}_g$, which is elliptic on $K$ (the injectivity of the symbol of $\mathcal{D}_g$ implies the bijectivity of the symbol of  $\mathcal{D}_g^\ast \circ \mathcal{D}_g$ by straightforward algebra).  Therefore $K$ should be viewed as the unknown quantity in this equation and $\phi$ should be viewed as a free datum.  Note that the correspondence between pairs $(K, \phi)$ and tensors $\chi$ is bijective.
 
Now consider the second and fourth equations of \eqref{eqn:ec}.  Again, in \cite{me3}, both these equations contain the operator $\div_g$ which has surjective symbol, making it under-determined elliptic.  One should thus imagine that each of these equations determines only \emph{part} of the unknown tensors $S$ and $\bar{S}$, say in some direct sum decomposition, while leaving the other part free.  This idea is implemented by using a modified version of the well-known \emph{York decomposition} of trace-free, symmetric tensors in the form of the mapping $\mathcal{Y}$.  This is needed to guarantee the uniqueness of the solutions of the extended constraint equations.  (The modification is that traces and divergences are being taken with respect to the fixed metric $\delta$ rather than the solution metric $g$ whenever the tensors in question correspond to free data.)   One can easily check that $\mathcal{Y}$ is a bijection.  Substituting this decomposition into the second and fourth equations of \eqref{eqn:ec} yields the second and fourth equations of the secondary system, and the operators $\div_{g} \circ \div_{g}^{\ast} (X)$ and $\div_{g} \circ \div_{g}^{\ast} (\bar{X})$ appearing there are elliptic on $X$ and $\bar{X}$ (again, the surjectivity of the symbol of $\div_g$ implies the bijectivity of the symbol of $\div_{g} \circ \div_{g}^{\ast}$).  Therefore $X$ and $\bar{X}$ should once again be viewed as the unknowns in these equations and $T$ and $\bar{T}$ can be treated as free data and the correspondence between $(S, \bar S)$ and $(X, \bar X, T, \bar T)$ is bijective.

Finally, consider the third and remaining equation of \eqref{eqn:ec}.  As pointed out in \cite{me3}, it is a well-known fact that the Ricci operator is not elliptic on $h$ because of its gauge invariance.  Again as in \cite{me3}, the standard trick of using harmonic coordinates will be employed to break the gauge invariance.  Recall that such coordinates $x^{a}$ are defined by the requirement that $\Delta_{g} x^{a} = 0$ for each $a$, making the $x^{a}$ harmonic functions.  It is well-known that under this coordinate condition, the Ricci operator can be written as in \eqref{eqn:redric} with $\Gamma_a = 0$ and that 
\begin{equation}
    R_{ab}^{H} = - \frac{1}{2} g^{rs} g_{ab,rs} + q(\Dif g) \, ,
    \label{defn:redric}
\end{equation}
where $q(\Dif g)$ denotes a term that is quadratic in the first derivatives of the components of $g$.  The operator  $\Ric^H$ is clearly elliptic when acting on $g$.   The third and remaining equation of the secondary system, then, is obtained simply by replacing the Ricci operator in \eqref{eqn:ec} by its formal expression in harmonic coordinates.

\section{Statement and Proof of Theorem A}

\subsection{The Statement}

The Implicit Function Theorem will be used to solve the secondary system near the trivial solution and so it is restated here for ease of reference.  For an excellent discussion and proof of this theorem, see
\cite{amr}.

\medskip \noindent \bfseries Implicit Function Theorem\mdseries:
\itshape Let $\Psi : \mathcal{X} \times \mathcal{B} \rightarrow
\mathcal{Y}$ be a smooth map between Banach spaces and suppose that
$\Psi(\bo ;  \bo) = \bo$.  If the restricted linearized operator $D \Psi(\bo ;  \bo)
\big|_{\mathcal{X} \times \{\bo\}} : \mathcal{X} \rightarrow
\mathcal{Y}$ is an isomorphism, then there exists an open set
$\mathcal{U} \subset \mathcal{B}$ containing $\bo$ and a smooth function
$\psi: \mathcal{U} \rightarrow \mathcal{X}$ with $\psi(\bo) = \bo$ so that $\Psi
\big( \psi(b), b \big) = \bo$.  \upshape \medskip

The Implicit Function Theorem allows solutions of the equation $\Psi(x,b) = \bo$, with $(x,b)$ sufficiently close to $(\bo; \bo)$ in the Banach space norm of $\mathcal{X} \times \mathcal{B}$ to be parametrized over the Banach space $\mathcal{B}$ of free data.  To apply this theorem to the secondary system, identify the three Banach spaces $\mathcal{X}$, $\mathcal{B}$ and $\mathcal{Y}$ with the space of unknown quantities $(h,K,X,\bar{X})$ appearing in the secondary system \eqref{eqn:assoc}, the space of free data $(T, \bar{T}, \phi)$ appearing there, and the space corresponding to the range of the secondary system, respectively.  That is, define
\begin{equation*}
    \begin{aligned}
	\mathcal{X} &= H^{k,\beta} \big( \sym \big) \times H^{k-1,
	\beta -1} \big( \symd \big) \times H^{k-1, \beta -1} \big(
	\oneform \big) \times H^{k-1, \beta -1} \big( \oneform \big) \\
	\mathcal{B} &= H^{k-2, \beta-2} \big( \symtt \big) \times
	H^{k-2, \beta-2} \big( \symtt \big) \times H^{k-1, \beta -1}(
	\R^{3}) \\
	\mathcal{Y} &= H^{k-2, \beta -2} \big( \symtf \big) \times
	H^{k-3, \beta -3} \big( \oneform \big) \times H^{k-2, \beta
	-2} \big( \sym \big) \times H^{k-3, \beta -3} \big( \oneform
	\big) \, .
    \end{aligned}
\end{equation*}
Now use the secondary system to define a map $\Psi :
\mathcal{X} \times \mathcal{B} \rightarrow \mathcal{Y}$ by
\begin{equation}
    \label{eqn:assocmap}
    \Psi(h, K, X, \bar{X}; T, \bar{T}, \phi) = \left(
    \begin{array}{c}
    \left[ \mathcal{D}_{g}^{\ast} \big( \mathcal{D}_{g} (\chi) - \star
    \, \bar{S} \big) \right]_{ab} \\
    \left[\div_{g} (\bar{S}) \right]_{a} - \eps^{st}_{ \hspace{1.5ex}
    a} \chi^{u}_{s} R_{ut}(g) \\
    \left[ \Ric^{H}(\delta + h) \right]_{ab} - S_{ab} + \chi^{s}_{s}
    \chi_{ab} - \chi^{s}_{a} \chi_{bs} \\
    \left[ \div_{g} (S) \right]_{a} + \chi^{st} \eps^{u}_{
    \hspace{1.5ex} at} \bar{S}_{su} \, .
    \end{array}
    \right)
\end{equation}
where $g = \delta + h$ and $S (X, T)$, $\bar{S}(\bar{X}, \bar{T})$ and $\chi(K, \phi)$ are
as in Definition \ref{defn:sec}.

The map $\Psi$ is a well-defined map of Banach spaces by the considerations of Section 2.3 (and the remark made there about non-linear terms) and is clearly smooth.  Since the flat solution $g = \delta$ and $\chi = S = \bar{S} = 0$ satisfies the extended constraint equations, $\Psi(\bo;\bo) = \bo$, and any other asymptotically flat solution of the secondary system satisfies $\Psi (h,K,X,\bar{X}; T, \bar{T}, \phi) = \bo$.   The first theorem to be proved in this paper is the following.
\begin{nonumthm}
    Let $\Psi : \mathcal{X} \times \mathcal{B} \rightarrow
    \mathcal{Y}$ be the map of Banach manifolds corresponding to the
    secondary system given above.  Then there is a neighbourhood
    $\mathcal{U}$ of zero in $\mathcal{B}$ and a smooth map of Banach
    manifolds $\psi : \mathcal{U} \rightarrow \mathcal{X}$ with
    $\psi(\bo) = \bo$ so that for each $b \in \mathcal{U}$, the
    point $(\psi(b), b)$ solves the secondary system \emph{up to a
    finite dimensional error}.  In other words, there is a
    finite-dimensional linear subspace $\mathcal{C} \subseteq
    \mathcal{Y}$ such that $\Psi(\psi(b), b) \in \mathcal{C}$. 
    Furthermore, the map $\psi$ is injective.
\end{nonumthm}

\subsection{The Proof} 

\paragraph*{Analysis of the linearization.} The first step of the proof is to show that the linearization of $\Psi$ in the $\mathcal{X}$ directions at the origin is Fredholm and to find its kernel and co-kernel.  This begins with a lemma.    
    
\begin{lemma} 
    \label{lemma:coker}
    The operator $\div_{\delta} \circ \div_{\delta}^{\ast} : H^{k-1,
    \beta - 1} \big( \oneform \big) \rightarrow H^{k-3, \beta - 3}
    \big( \oneform \big)$ is Fredholm and injective with image equal to the
    codimension-3 space
    $$\mathcal{I}_{1} = \left\{ X = X_{a} \dif x^{a} \in H^{k-3, \beta
    - 3} \big( \oneform \big) \, : \, \int_{\R^{3}} X_{a} = 0
    \: \forall \, a \right\} \, .$$
    Furthermore, the operator $\mathcal{D}_{\delta}^{\ast} \circ
    \mathcal{D}_{\delta} : H^{k-1, \beta -1} \big( \symd \big)
    \rightarrow H^{k-3, \beta - 3} \big( \symd \big)$ is Fredholm and injective
    with image equal to the codimension-5 space
    $$\mathcal{I}_{2} = \left\{ T = T_{ij} \dif x^{i} \oplus \dif
    x^{j} \in H^{k-3, \beta - 3} \big( \symd \big) \, : \, \int_{R^{3}}
    T_{ij} = 0 \: \forall \, i,j \right\} \, .$$
    Here, $x^{i}$ are the standard coordinate functions on $\R^{3}$.
\end{lemma}
    
\begin{proof} 
    
This lemma follows from Proposition \ref{prop:inv} on the mapping properties of the linear, homogeneous elliptic operators with constant coefficients, since $\div_{\delta} \circ \div_{\delta}^{\ast}$ and $\mathcal{D}_{\delta} ^{\ast} \circ \mathcal{D}_{\delta}$ are such operators.  It is necessary only to identify the kernels of these operators and the kernels of their adjoints.  Note that this task is simplified somewhat, since these operators are formally self-adjoint.
    
Let $P$ be any constant-coefficient elliptic operator on $\R^3$.  By Proposition \ref{prop:inv} the components of a solution of $Pu = 0$ with growth or decay at infinity are polynomials of degrees less than or equal to the growth rate, or zero, respectively.  The operators considered by the lemma are thus injective since $\beta \in (-1,0)$.  Furthermore, $\beta \in (-1,0)$ implies that $-\beta -1 \in (0,1)$ and so the tensors in the kernels of the adjoints of these operators have components that are polynomials of degree zero.  That is, these components must all be constants.  Consequently,  the images of the operators considered in the lemma are tensors orthogonal to the constant 1-forms $\lambda_{i} \dif x^{i}$ ($\lambda_{i} \in \R$ for all $i$) and the constant trace-free symmetric tensors $\mu_{ij} \dif x^{i} \oplus \dif x^{j}$ ($\mu_{ij} \in \R$ for each $i,j$ and $\delta^{ij} \mu_{ij} = 0$), respectively.  The dimension of each of these spaces is obviously 3 and 5, respectively.
\end{proof}

\begin{prop}
    \label{prop:lin}
    The linearized operator $P \equiv \Dif \Psi(\bo ; \bo) \big|_{\mathcal{X} \times \{\bo\}}$ in the $\mathcal{X}$     
    directions is given by
    \begin{equation}
	\label{eqn:lin}
	P(h,K,X,\bar{X}) = \left(
	\begin{array}{c}
	    \mathcal{D}_{\delta}^{\ast} \big( \mathcal{D}_{\delta} (K)
	    - \star \, \div_{\delta}^{\ast} (\bar{X}) \big)
	    \\[0.75ex]%
	    \div_{\delta} \circ \div_{\delta}^{\ast} (\bar{X})
	    \\[0.75ex]%
	    -\frac{1}{2} \Delta_{\delta} h - \div_{\delta}^{\ast} (X)
	    \\[0.75ex]%
	    \div_{\delta} \circ \div_{\delta}^{\ast} (X)
	\end{array}
	\right)
    \end{equation}
where $\Delta_{\delta}$ is the Laplacian corresponding to the
Euclidean metric.  The principal symbol of $P$ is bijective, making
$P$ an elliptic operator of Banach spaces.  Furthermore, $P$ is Fredholm and
injective with image equal to the codimension-11 space
$$\mathrm{Im}(P) = \mathcal{I}_{2} \times \mathcal{I}_{1} \times
H^{k-2,\beta-2} \big( \sym \big) \times \mathcal{I}_{1} \, ,$$
where $\mathcal{I}_{1}$ and $\mathcal{I}_{2}$ are the spaces defined
in Lemma \ref{lemma:coker}.
\end{prop}

\begin{proof}
    
One can break up the calculation of the linearization into two pieces:
\begin{equation*}
    P(h,K,X,\bar{X}) =  \Dif \Psi(\bo; \bo)(h,0,0,0; \bo) + \Dif \Psi(\bo; \bo) (0,K,X,\bar{X}; \bo) \, .
\end{equation*}
First,
\begin{equation}
        \Dif \Psi(\bo; \bo) (0,K,X,\bar{X}; \bo) =  \left(
        \begin{array}{c}
	   \mathcal{D}_{\delta}^{\ast} \big( \mathcal{D}_{\delta} (K)
	   - \star \, \div_{\delta}^{\ast} (\bar{X}) \big)
	   \\[0.75ex]
        	   \div_{\delta} \circ \div_{\delta}^{\ast} (\bar{X})
	   \\[0.75ex]%
	   - \div_{\delta}^{\ast} (X)
	   \\[0.75ex]%
	   \div_{\delta} \circ \div_{\delta}^{\ast} (X)
        \end{array}
    \right)
    \label{eqn:linone}
\end{equation}    
since $\Psi(0, K,X,\bar{X}; \bo)$ consists of a sum of differential
operators that are linear in $K,X$ and $\bar{X}$ with terms which are
quadratic in $K,X$ and $\bar{X}$.  Second,
\begin{equation}
    \Dif \Psi(\bo; \bo)(h,0,0,0; \bo) =
    \left.  \frac{\dif}{\dif s} \left(
    \begin{array}{c}
	0\\
	0\\
	\Ric^{H}(\delta + sh) \\
	0
    \end{array}
    \right) \right|_{s=0} 
    = \left(
    \begin{array}{c}
	0\\
	0\\
	-\frac{1}{2} \Delta_{\delta} h \\
	0
    \end{array} 
    \right)
    \label{eqn:lintwo}
\end{equation}
by definition of the reduced Ricci operator and using the fact that
the terms in $\Gamma$ appearing there are quadratic. 
    
Since the operator $P$ is upper-triangular and the operators appearing on the diagonal are all elliptic, $P$ is itself elliptic.  Furthermore, Proposition \ref{prop:inv} shows that $P$ is Fredholm; and by the choice of $\beta \in (-1,0)$ made in Section 2.3, both $\beta$ and $\beta - 1$ are less than 0, so that each of the operators appearing on the diagonal are injective on their respective domains.  The operator $P$ is thus itself injective.

To find the image of $P$, one must attempt to solve the equations
\begin{equation}
    \left(
    \begin{array}{c}
	\mathcal{D}_{\delta}^{\ast} \big( \mathcal{D}_{\delta} (K)
	- \star \, \div_{\delta}^{\ast} (\bar{X}) \big)
	\\[0.75ex]%
	\div_{\delta} \circ \div_{\delta}^{\ast} (\bar{X})
	\\[0.75ex]%
	-\frac{1}{2} \Delta_{\delta} h - \div_{\delta}^{\ast} (X)
	\\[0.75ex]%
	\div_{\delta} \circ \div_{\delta}^{\ast} (X)
    \end{array}
    \right) = \left(
    \begin{array}{c}
	f_{1} \\ f_{2} \\ f_{3} \\ f_{4}
    \end{array}
    \right)
    \label{eqn:solveme}
\end{equation}
with $(f_{1}, f_{2}, f_{3}, f_{4}) \in \mathcal{Y}$.  According to Lemma \ref{lemma:coker}, the second and fourth equations can be solved if and only if $f_{2}$ and $f_{4}$ are in $\mathcal{I}_{1}$.  The 
third equation can now be solved since $\Delta_{\delta}$ is an isomorphism according to Proposition \ref{prop:inv}.  The remaining equation of \eqref{eqn:solveme} can be solved if and only if
$$\int_{\R^{3}} [f_{1}]_{ij} + [ \mathcal{D}_{\delta}^{\ast} \star
\div_{\delta}^{\ast}(\bar{X}) ]_{ij} = 0$$
for every $i$ and $j$.  But this is true if and only if $f_{1} \in
\mathcal{I}_{2}$ by the definition of the adjoint and the fact that
constants are in the kernel of $\mathcal{D}_{\delta}$.  Thus $(f_{1},
f_{2}, f_{3}, f_{4}) \in \mathcal{I}_{2} \times \mathcal{I}_{1} \times
H^{k-2,\beta-2} \big( \sym \big) \times \mathcal{I}_{1}$, which is of codimension 11 in $\mathcal{Y}$.
\end{proof}
 
\paragraph*{Existence of Solutions in Theorem A.}  The conclusion to be drawn from Proposition \ref{prop:lin} is that solutions of equation $\Psi(x; b) = \bo$ can not be found using the Implicit Function Theorem for $x = (h,K,X,\bar{X})$ near $\bo$ in terms of $b = (T, \bar{T}, \phi)$.  The non-trivial co-kernel of the linearization $P$ is the essential obstruction.  However, the existence of solutions up to an \emph{error term} transverse to the image of $P$ can be proved using the following 
technique. 

Since $P$ is Fredholm with 11-dimensional co-kernel, the image of $P$ is closed and one can find (in many different ways) an 11-dimensional subspace $\mathcal{C}$ so that
$$\mathcal{Y} = \mathrm{Im}(P) \oplus \mathcal{C} \, .$$
If $\eta : \R^{3} \rightarrow \R$ denotes a smooth, positive function on $\R^{3}$ with compact support satisfying $\int_{\R^{3}} \eta = 1$, then one such choice is given by
\begin{equation*}
    \mathcal{C} = \left\{\eta \mu_{ij} \dif x^{i} \otimes \dif x^{j}
    \, : \, \mu_{ij} \in \R \right\} \times \left\{ \eta
    \bar{\lambda}_{i} \dif x^{i} \, : \, \bar{\lambda}_{i} \in \R
    \right\} \times \{0\} \times \left\{ \eta \lambda_{i} \dif x^{i}
    \, : \, \lambda_{i} \in \R \right\}
\end{equation*}
If $\pi : \mathcal{Y} \rightarrow \mathrm{Im}(P)$ denotes the projection operator corresponding to this decomposition, namely the operator given by
\begin{equation*}
    \pi(f_{1}, f_{2}, f_{3}, f_{4}) = \left(
    \begin{array}{c}
	f_{1} - \eta \left( \int_{\R^{3}} [f_{1}]_{ij} \right) \dif
	x^{i} \otimes \dif x^{j} \\[0.5ex]%
	f_{2} - \eta \left( \int_{\R^{3}} [f_{2}]_{i} \right) \dif
	x^{i} \\[0.5ex]%
	f_{3} \\[0.5ex]%
	f_{4} - \eta \left( \int_{\R^{3}} [f_{4}]_{i} \right) \dif
	x^{i}
    \end{array}
    \right) \, ,
\end{equation*}
then the operator $\pi \circ \Psi : \mathcal{X} \times \mathcal{B} \rightarrow \mathrm{Im}(P)$ has linearization in the $\mathcal{X}$ direction equal to $\pi \circ P$, which is injective (proved as in
Proposition \ref{prop:lin}) and is surjective since the composition with $\pi$ forces it to map onto its image.  Consequently, the Implicit Function Theorem can be applied to the equation $\pi \circ
\Psi (x,b) = \bo$.  The result is as follows.

There is a neighbourhood $\mathcal{U}$ of $\bo$ in $\mathcal{B}$ and a smooth function $\psi : \mathcal{B} \rightarrow \mathcal{X}$ such that if $b = (T, \bar{T}, \phi) \in \mathcal{U}$ and $\psi(b) \equiv \big( h(b), K(b), X(b), \bar{X}(b) \big)$, then $\big( \psi(b), b \big) \in \mathcal{X} \times \mathcal{B}$ satisfies the equation $\pi \circ \Psi \big( \psi(b), b \big) = \bo$.  This is equivalent to the statement that the tensors
\begin{align*}
    &g = \delta + h \\
    &\chi = K + \tfrac{1}{3}\phi g \\
    &S = \div_{g}^{\ast} \big( X \big) + T - \tfrac{1}{3} Tr_{g}(T)
    g \\
    &\bar{S} = \div_{g}^{\ast} \big( \bar{X} \big) + \bar{T} -
    \tfrac{1}{3} Tr_{g}(\bar{T}) g
\end{align*}
satisfy the equations
\begin{equation}
    \label{eqn:thmone}
    \left(
    \begin{array}{c}
    \left[ \mathcal{D}_{g}^{\ast} \big( \mathcal{D}_{g} (\chi) - \star
    \, \bar{S} \big) \right]_{ab} \\[0.5ex]%
    \left[\div_{g} (\bar{S}) \right]_{a} - \eps^{st}_{ \hspace{1.5ex}
    a} \chi^{u}_{s} R_{ut}(g) \\[0.5ex]%
    \left[ \Ric^{H}(\delta + h) \right]_{ab} - S_{ab} + \chi^{s}_{s}
    \chi_{ab} - \chi^{s}_{a} \chi_{bs} \\[0.5ex]%
    \left[ \div_{g} (S) \right]_{a} + \chi^{st} \eps^{u}_{
    \hspace{1.5ex} at} \bar{S}_{su} 
    \end{array}
    \right) = \eta \left( 
    \begin{array}{c} 
	\mu_{ab} \\[0.5ex]%
	\bar{\lambda}_{a}\\[0.5ex]%
	0\\[0.5ex]%
	\lambda_{a}
    \end{array}
    \right) \in \mathcal{C} \, ,
\end{equation}
where
\begin{align*}
    \mu_{ab} &= \int_{\R^{3}} \left[ \mathcal{D}_{g}^{\ast} \big(
    \mathcal{D}_{g} (\chi) - \star \, \bar{S} \big) \right]_{ab}
    \\[0.5ex]%
    \bar{\lambda}_{a} &= \int_{\R^{3}} \left( \left[\div_{g} (\bar{S})
    \right]_{a} - \eps^{st}_{ \hspace{1.5ex} a} \chi^{u}_{s} R_{ut}(g)
    \right) \\[0.5ex]%
    \lambda_{a} &= \int_{\R^{3}} \left( \left[ \div_{g} (S)
    \right]_{a} - \chi^{st} \eps^{u}_{ \hspace{1.5ex} at}
    \bar{S}_{su}\right) \, ,
\end{align*}
in the standard coordinates of $\R^{3}$, provided that the free data
$(T, \bar{T}, \phi)$ is chosen sufficiently small in the norm of the
Banach space $\mathcal{B}$.  
 
\paragraph*{Uniqueness of Solutions in Theorem A.} It remains to show that the mapping from $(T, \bar{T}, \phi)$ to the solution $(h, K, X, \bar{X})$ is injective.  This will follow from the Implicit Function Theorem if it can be shown that the linearization of $\Psi$ in the direction of the free data, namely $\Dif(\pi \circ \Psi(\bo; \bo)) \big|_{\{0\} \times \mathcal{B}}$, is injective.  But a calculation similar to that in Proposition \ref{prop:lin} gives
\begin{align}
    \label{eqn:linb}
    \Dif \big( \pi \circ \Psi(\bo;\bo) \big) \big|_{\{0\} \times \mathcal{B}} ( T, \bar{T}, \phi) &= \left. 
    \frac{\dif}{\dif s} \right|_{s=0} \pi \circ \Psi(0, 0, 0, 0; sT, s\bar{T},
    s\phi) \notag \\
    &= \left.  \frac{\dif}{\dif s} \right|_{s=0} s \left(
    \begin{array}{c}
	\left[ \mathcal{D}_{\delta}^{\ast} \big( \mathcal{D}_{\delta}
	\big( \tfrac{1}{3} \phi \delta \big) - \star \, \big(\bar{T} -
	\tfrac{1}{3} Tr_{\delta}(\bar{T}) \delta \big) \big)
	\right]_{ab} \\[0.25ex]
    	\left[\div_{\delta} \big(\bar{T} - \tfrac{1}{3}
	Tr_{\delta}(\bar{T}) \delta \big) \right]_{a} \\[0.25ex]
	\left[- T + \tfrac{1}{3} Tr_{\delta}(T) \delta \right]_{ab}\\[0.25ex]
	\left[ \div_{\delta} \big (T - \tfrac{1}{3} Tr_{\delta}(T)
	\delta \big) \right]_{a}
    \end{array}
    \right) \notag \\
    &\qquad - \left. \frac{\dif}{\dif s} \right|_{s=0} s \left(
         \begin{array}{c}
	     \eta \int_{\R^3}\left[ \mathcal{D}_{\delta}^{\ast} \big( \mathcal{D}_{\delta}
	     \big( \tfrac{1}{3} \phi \delta \big) - \star \, \big(\bar{T} -
	     \tfrac{1}{3} Tr_{\delta}(\bar{T}) \delta \big) \big)
	     \right]_{ab} \\[0.25ex]
	     \eta \int_{\R^3}\left[\div_{\delta} \big(\bar{T} - \tfrac{1}{3}
	     Tr_{\delta}(\bar{T}) \delta \big) \right]_{a} \\[0.25ex]
	     0\\[0.25ex]
	     \eta \int_{\R^3}\left[ \div_{\delta} \big (T - \tfrac{1}{3} Tr_{\delta}(T)
	     \delta \big) \right]_{a}
         \end{array}
         \right) \notag \\
    &= \left(
    \begin{array}{c}
	\nabla_{a} \nabla_{b} \phi - \eps^{eb}_{\hspace{1.5ex} c}
	\nabla_{b} \bar{T}_{ae} - \eps^{eb}_{\hspace{1.5ex} c}
	\nabla_{b} \bar{T}_{ce} \\ - \eta \int_{\R^{3}} \big(
	\nabla_{a} \nabla_{b} \phi - \eps^{eb}_{\hspace{1.5ex} c}
	\nabla_{b} T_{ae} - \eps^{eb}_{\hspace{1.5ex} c} \nabla_{b}
	T_{ce} \big)\\[1ex]
	0 \\
	- T_{ab}\\
	0
    \end{array}
    \right) 
\end{align}
using the fact that $T$ and $\bar{T}$ were chosen to be
transverse-traceless with respect to $\delta$.  
         
Suppose now that $\Dif (\pi \circ \Psi) \big|_{\{0\} \times \mathcal{B}} (\bo ; \bo)(T, \bar{T}, \phi) = (0,0,0,0)$.  Then clearly $T = 0$.  Taking the trace of what remains  of the first equation in \eqref{eqn:linb} yields
$$ \Delta_{\delta} \phi - \eta \int_{\R^{3}} \Delta_{\delta} \phi = 
\Delta_{\delta} \phi = 0$$
by the divergence theorem (valid because of the decay property of $\phi$).  But now, one can invoke the injectivity of $\Delta$ on the space $H^{k-1, \beta-1}(\R^{3})$ to conclude that $\phi = 0$. 
Finally, use the identity $\eps^{e}_{\hspace{1ex} bc } \bar{T}_{ae} + \eps^{e}_{\hspace{1ex} ab} \bar{T}_{ce} + \eps^{e}_ {\hspace{1ex} ca} \bar{T}_{be} = 0$ and the fact that $\bar{T}$ is divergence free at the Euclidean metric to conclude that the remaining equation for $\bar{T}$ reads
$$\eps^{eb}_{\hspace{1.5ex} a} \nabla_{b} \bar{T}_{ce} - \eta
\int_{\R^{3}} \eps^{eb}_{\hspace{1.5ex} a}  \nabla_{b} \bar{T}_{ce} = 0 \, .$$
The integrand above is an exact vector-valued differential.  Thus by using Stokes' Theorem (valid because of the decay property of $\bar{T}$), one can eliminate the integral above and end up with
$$\eps^{eb}_{\hspace{1.5ex} a}  \nabla_{b} \bar{T}_{ce} = 0 \, .$$
This equation shows that the vector-valued one-form $\bar{T}_{ab} \dif x^b$ is closed and hence exact, so that $\bar{T}_{ab} = \nabla_{b} \bar{T}_{a}$ for some vector-valued zero form $\bar{T}_{a}$.  The divergence-free condition on $\bar{T}_{ab}$ then shows that $\Delta_{\delta} \bar{T}_{b} = 0$ and so $\bar{T}_b = 0$ by the injectivity of $\Delta_{\delta}$ on the space of $H^{k-2, \beta - 2}$ tensors over $\R^{3}$.

These calculations show that the operator $\Dif ( \pi \circ \Psi) \big|_{\{ 0 \} \times \mathcal{B}}(\bo ; \bo)$ is injective, thereby proving the uniqueness claim contained in Theorem A and concluding the proof of Theorem A. \hfill \QED

\section{Statement and Proof of Theorem B}

\subsection{The Statement}

In order to complete the proof of the Main Theorem, it remains to show that the solution \eqref{eqn:thmone} of the secondary system constructed in the previous section actually satisfies the extended
constraint equations.  The second theorem to be proved in this paper establishes this fact. 

\begin{nonumthm}
    If $(\psi(b),b)$ is a solution of the equation $\pi \circ
    \Psi(\psi(b),b) = \bo$ as in Theorem A, with $b= (T, \bar{T},
    \phi)$ sufficiently small in the norm of the Banach space
    $\mathcal{B}$ and $\psi(b) = (h(b), K(b), X(b), \bar{X}(b))$, then
    the quantities
    \begin{align*}
	&g = \delta + h(b) \\
	&\chi = K(b) + \tfrac{1}{3}\phi \delta \\
	&S = \div_{g}^{\ast} \big( X(b) \big) + T - \tfrac{1}{3} Tr_{g}(T)
	g \\
	&\bar{S} = \div_{g}^{\ast} \big( \bar{X}(b) \big) + \bar{T} -
	\tfrac{1}{3} Tr_{g}(\bar{T}) g \, ,
    \end{align*}
    satisfy the extended constraint equations \eqref{eqn:ec}.
\end{nonumthm}

\paragraph*{Strategy of Proof.}  Theorem A has produced 
\begin{equation*}
    \begin{aligned}
	g &\in H^{k, \beta}(\sym)\qquad \\ 
	\chi &\in H^{k-1, \beta-1}(\sym) \\
	S &\in H^{k-2, \beta-2}(\symtf)\qquad \\
	 \bar{S} &\in H^{k-2, \beta-2}(\jac) 
    \end{aligned}
\end{equation*}
and error terms $\lambda_a$, $\bar{\lambda}_a$ and $\mu_{ab}$ in $\R$ that satisfy equation
\eqref{eqn:thmone}.  Let $J\in H^{k-2, \beta-2}(\jac)$ be the tensor $J = \mathcal{D}_{g}(\chi) - \star \bar{S}$.  Recall that $(g, \chi, S, \bar{S})$ satisfies the extended constraint equations \eqref{eqn:ec} if and only if
\begin{equation}
    \label{eqn:zeroqty}
    \begin{aligned}
	J_{abc} &= 0 \\
	\Gamma^a &= 0 \\
	\mbox{error terms} &= 0
    \end{aligned}
\end{equation}
where $\Gamma^a$ is the quantity arising from the harmonic gauge choice as explained in Section 2.2.
	
Here is a schematic of how it will be shown that \eqref{eqn:zeroqty} holds.  Let $z_{\eps} = (J, \Gamma)$ represent a solution of the secondary system, where the norm of the free data is represented by the small parameter $\eps$, and let $\Lambda_{\eps}$ represent an error term lying in a fixed, finite-dimensional subspace $\mathcal{C}$.  The next proposition will use the as-yet-unexploited structure --- the integrability conditions --- contained within the extended constraint equations to show that $z_{\eps}$ satisfies a system of equations of the form $$\mathcal{P}_ {\eps}(z_{\eps}) = \Lambda_{\eps} \, ,$$  where $\mathcal{P}_{\eps}$ is a family of linear, elliptic operators with coefficients depending smoothly on $\eps$.  These equations will be called the \emph{auxiliary equations}.  Then, it will be shown that the operator $\mathcal{P}_{\eps}$ is (1) injective and uniformly elliptic, and thus uniformly injective, all for sufficiently small $\eps$, and (2) that the finite-dimensional subspace $\mathcal{C}$ is transverse to the image of $\mathcal{P}_0$.  Consequently, the stability of co-kernels of linear elliptic operators proved in Proposition \ref{prop:stab} shows that $\mathcal{C}$ remains transverse to the image of $\mathcal{P}_{\eps}$ for sufficiently small $\eps$.  Since $\Lambda_{\eps} \in \mathcal{C} \cap \mathrm{Im} (\mathcal{P}_{\eps})$, therefore $\Lambda_{\eps} = 0$.  But now the uniform injectivity can be invoked yet again to show that $z_\eps = 0$.

\subsection{The Proof}

\paragraph*{The Auxiliary Equations.}  The first step in the proof of Theorem B is to derive the auxiliary equations satisfied by $J$ and $\Gamma$ which are a consequence of the integrability conditions in the conformal constraint equations.

\begin{prop}
    The quantities $J$, $\Gamma$ and the real numbers $\mu_{ab}$, 
    $\lambda_{a}$, $\bar{\lambda}_{a}$ satisfy the equations
    \begin{equation}
	\label{eqn:gam}
	\frac{1}{2} \big( \Delta_{g} \Gamma_{b} - R^{a}_{b} \Gamma_{a}
	\big) = - \eta \lambda_{b} - \chi^{st} J_{sbt} + A^{a}
	\big(\chi_{ab} - \chi^{s}_{s} g_{ab} \big) \, .
    \end{equation}
    where $A_{b} = g^{ac} J_{abc}$, and   
    \begin{equation}
	\label{eqn:j}
	\begin{aligned}
	    \big[ \mathcal{D}_{g}^{\ast} (J) \big]_{ab} &= \eta
	    \mu_{ab} \\
	    \big[ \mathcal{Q}_{g} (J) \big]_{a} &= 2 \eta
	    \bar{\lambda}_{a} \, ,
	\end{aligned}
    \end{equation}
    where $\mathcal{Q}_{g}$ is the operator defined at the beginning
    of section 2.1.
\end{prop}

\begin{proof} 
    
The equation satisfied by $\Gamma$ will be deduced from the Bianchi identity $\nabla^{a} R_{ab} - \frac{1}{2} \nabla_{b} R = 0$.  In fact, substitute the reduced Ricci operator into this identity to obtain
\begin{align}
    0 &= \nabla^{a} R^{H}_{ab} - \frac{1}{2} \nabla_{b} R^{H} +
    \frac{1}{2} \big( \Delta_{g} \Gamma_{b} + \nabla^{a} \nabla_{b}
    \Gamma_{a} - \nabla_{b} \nabla^{a} \Gamma_{a} \big) \notag \\
    &= \nabla^{a} S_{ab} - \nabla^{a} (\chi^{s}_{s} \chi_{ab}) +
    \nabla^{a} (\chi^{s}_{a} \chi_{sb}) + \frac{1}{2} \big( \nabla_{b}
    (\chi^{s}_{s})^{2} - \chi^{st} \chi_{st} \big) + \frac{1}{2} 
    \big( \Delta_{g} \Gamma_{b} + R^{a}_{b} \Gamma_{a} \big) \notag \\
    &= \eta \lambda_{b} - \chi^{st} \eps^{e}_{\hspace{0.5ex} bt}
    \bar{S}_{se} + \frac{1}{2} \big( \Delta_{g} \Gamma_{b} + R^{a}_{b}
    \Gamma_{a} \big) \notag \\
    &\qquad - \big( \nabla^{s} \chi^{a}_{s} - \nabla^{a} \chi^{s}_{s}
    \big) \big( g_{ab} \chi^{s}_{s} - \chi_{ab} \big) - \big(
    \nabla_{b} \chi_{st} - \nabla_{s} \chi_{b} \big) \chi^{st} \, .
    \label{eqn:gamone}
\end{align}
Now let $A_{b} = g^{ac} J_{abc}$.  According to the definition of $J$ made above, $A_{b} = \nabla^{a} \chi_{ab} - \nabla_{b} \chi^{a}_{a}$.  Substituting this into \eqref{eqn:gamone} leads to
\begin{equation*}
    \frac{1}{2} \big( \Delta_{g} \Gamma_{b} + R^{a}_{b} \Gamma_{a}
    \big) = - \eta \lambda_{b} - \chi^{st} J_{sbt} + A^{a}
    \big(\chi_{ab} - \chi^{s}_{s} g_{ab} \big) \, ,
\end{equation*}
which is the desired equation for $\Gamma_{a}$.
    
Next, the first equation of \eqref{eqn:j} holds for $J$ by definition.  For the second equation, consider
\begin{align}
    \big[ \mathcal{Q}_{g} (J) \big]_{a} &= \eps^{ebc} \nabla_{e} 
    J_{abc} \notag \\
    &= \eps^{ebc} \nabla_{e} \big( ( \nabla_{c} \chi_{ab} - 
    \nabla_{b} \chi_{ac}) - \eps^{s}_{\hspace{1ex} bc} \bar{S}_{as} 
    \big) \, . \notag \\
    \intertext{At this stage, one must commute the second derivatives, bringing in curvature terms.  Thus,}
    \big[ \mathcal{Q}_{g} (J) \big]_{a} &= \big( R_{ecas} \chi^{s}_{b} + R_{ecbs} \chi^{s}_{a} \big) 
    \eps^{ebc} - \eps^{ebc} \eps^{s}_{\hspace{1ex} bc} \nabla_{e} 
    \bar{S}_{as} \notag \\
    &= 2 R_{ecas} \chi^{s}_{b} \eps^{ecb} - 2 \nabla^{s} \bar{S}_{as}
    \label{eqn:jone}
\end{align}
by the algebraic Bianchi identity satisfied by the curvature tensor. Substitute the second equation of \eqref{eqn:thmone} into \eqref{eqn:jone} to obtain
\begin{equation*}
    \big[ \mathcal{Q}_{g} (J) \big]_{a} = 2 R_{ecas} \chi^{s}_{b}
    \eps^{ecb} - 2 \eps^{st}_{\hspace{1.5ex} a} \chi^{u}_{s} R_{ut} + 
    2 \eta \bar{\lambda}_{a} \, .
\end{equation*}
The fact that only curvature terms remain is a manifestation of the integrability conditions.  Finally, use the decomposition $R_{ecas} = R_{ea} g_{cs} - R_{es} g_{ca} + R_{cs} g_{ea} - R_{ca} 
g_{es} - \frac{1}{2} R \left( g_{ea} g_{cs} - g_{es} g_{ca} \right)$
to find that $R_{ecas} \chi^{s}_{b} \eps^{ecb} = \eps^{st}_{
\hspace{1.5ex} a} \chi^{u}_{s} R_{ut}$ and thus that 
\begin{equation*}
    \big[ \mathcal{Q}_{g} (J) \big]_{a} = 2 \eta \bar{\lambda}_{a} \, ,
\end{equation*}
which is the second equation of \eqref{eqn:j}. 
\end{proof}

\noindent \scshape Notation: \upshape denote the pair of operators appearing in \eqref{eqn:j} by $$\mathcal{P}_{\eps} (J) = \big( \mathcal{D}_g^\ast(J), \mathcal{Q}_g(J) \big) \, .$$ Thus $\mathcal{P}_\eps$ is an operator from $H^{k-2, \beta-2} \big( \jac \big)$ to $H^{k-3, \beta-3} \big( \symtf \times \oneform \big)$.

\paragraph*{Uniform Injectivity and Ellipticity.} The operators appearing on the left hand sides of equations \eqref{eqn:gam} and \eqref{eqn:j}  should be viewed as linear, elliptic operators for $J$ and $\Gamma$ whose coefficients depend on the solution $g$, $\chi$, $S$ and $\bar{S}$ of Theorem A and thus on the free data $(T, \bar{T}, \phi)$.   The uniform injectivity and ellipticity of these operators will be established one at a time using the following lemma.
  
\begin{lemma}
	\label{lemma:contra}    
	Let $g = \delta + h$ be an asymptotically flat metric on $\R^{3}$ with $h \in H^{k,\beta}\big( \sym \big)$ and let $\mathcal{R}(x)$ be a quantity that is quadratic in the argument $x$ with coefficients proportional to the curvature of $g$.  For any $l \geq 1$ and $\gamma < -1$, there exists a number $\eps > 0$ so that if $\Vert h \Vert_{H^{k,\beta}} \leq \eps$, then the only function $u \in H^{l,\gamma}(\R^{3})$ satisfying the inequality
	\begin{equation}
		\label{eqn:ineq}
		0 \leq - \int_{\R^{3}} \Vert \nabla u \Vert^{2} \dif \vol + C
		\int_{\R^{3}} \mathscr{R}(u) \dif \vol
    	\end{equation}
for some constant $C$, where $\mathscr{R}(u)$ is a quantity quadratic in its arguments and with coefficients proportional to the curvature of $g$, is the zero function.
\end{lemma}

\noindent \scshape Remark: \upshape This lemma is the generalization of a result from \cite{me3}.

\begin{proof}
    
Because $h \in H^{k,\beta}$ with $k \geq 4 >\frac{3}{2}$ and $\beta <0$, the Sobolev Embedding Theorem \cite{bartnik} guarantees that $\vert h \vert_{C^2} \leq C \Vert h \Vert_{H^{k,\beta}}$.  Consequently, the coefficients of $\mathcal{R}$ are bounded in $C^0$ by $\Vert h \Vert_{H^{k,\beta}}$ and it follows that
\begin{align}
    \int_{\R^3} \mathcal{R}(u) \dif \vol &\leq \int_{\R^{3}} \Vert \mathscr{R} \Vert \, \vert u \vert^{2} \dif \vol \notag \\
    &\leq C \Vert h \Vert_{H^{k,\beta}}
    \int_{\R^{3}} \vert u \vert^{2} \sigma^{-2} \dif \vol \notag  \\
    &\leq C \Vert h \Vert_{H^{k,\beta}} \int_{\R^{3}}\Vert \nabla u \Vert^{2} \dif \vol
    \label{eqn:poincare}
\end{align}
by the Poincar\'e inequality for weighted Sobolev norms proved in  \cite{bartnik}. (Actually, \cite{bartnik} proves this only for the metric $g = \delta$, but the result is also valid for the metric $g = \delta + h$ with sufficiently small $\Vert h \Vert_{H^{k,\beta}}$ because the norms with the metric $g$ are uniformly equivalent to the norms with the metric $h$ and no Christoffel symbols appear.) Using \eqref{eqn:poincare} in inequality \eqref{eqn:ineq} leads to 
\begin{equation*}
    0 \leq \left( C \Vert h \Vert_{H^{k,\beta}} - 1 \right)
    \int_{\R^{3}} \Vert \nabla u \Vert^{2} \, ,
\end{equation*}
so if $\Vert h \Vert_{H^{k,\beta}}$ is sufficiently small, the right hand side above becomes negative.  Avoiding this contradiction requires $\nabla u = 0$.  But since $\Vert u \Vert$ decays at infinity when $\gamma <-1$, it must be true that $u = 0$.
\end{proof}

It is now straightforward to deduce the uniform injectivity and ellipticity for the operator $\Delta_g - \Ric$ appearing in \eqref{eqn:gam}.  In fact, all that is needed is to show the injectivity of this operator, since uniform ellipticity is clear.  To see this, contract the equation $\Delta_g \Gamma_a - R^b_a \Gamma_b = 0$ with $\Gamma^{a}$ to obtain
$$0 = \Delta \Vert \Gamma \Vert^{2} - \Vert \nabla \Gamma \Vert - 2 Ric (\Gamma, \Gamma) \, .
$$
Then integrate over $\R^{3}$ and apply the divergence theorem as well as the Cauchy-Schwarz inequality to produce an expression of the form considered in Lemma \ref{lemma:contra}.  The injectivity of $\Delta_g - \Ric$ follows.

The analogous result for the operator $\mathcal{P}_\eps$ appearing in \eqref{eqn:j} is not quite so straightforward.  Another lemma is needed; then a lengthy calculation will transform the injectivity equation for $\mathcal{P}_\eps$ into a form suitable for the application of Lemma \ref{lemma:contra}.

\begin{lemma}
	\label{lemma:jacdecomp}
	Let $M$ be any Riemannian manifold.  Then $\mathcal{J} (M)$ is canonically isomorphic to $\mathcal{S}^2_{g}(M) \times \Lambda^{1} (M)$.  In local coordinates, this isomorphism is given by $J_{abc} = \frac{1}{2} \left( \eps^{e}_{\hspace{1ex} bc} F_{ae} +  A_{b} g_{ac} - A_{c} g_{ab} \right)$, where $\eps_{abc}$ is the fully antisymmetric permutation symbol associated to the metric of $M$.
\end{lemma}
    
\begin{proof} Given any tensor $J \in \mathcal{J}(M)$, one can define a 2-tensor $T$ by the prescription
$$J(X,Y,Z) = T \big( X, \big(\star \, (Y^{\flat} \wedge Z^{\flat}) \big)^{\sharp} \big) \, ,$$
for any three vectors $X$, $Y$, and $Z$, which is well-defined and unique because of the antisymmetry in the last two slots of $J$.  Here, $\sharp$ and $\flat$ are the raising and lowering operators associated to the metric of $M$ and $\star$ is its Hodge star operator.  Moreover, one can check that the trace of $T$ vanishes by virtue of the symmetries of $J$.  The tensor $T$ can now be decomposed into its symmetric part $F$ and its antisymmetric part written in terms of a 1-form $A$ by means of Hodge duality.  It is easy to see that this association is bijective and yields the
local coordinate expression of the lemma.
\end{proof}

\begin{prop}
    \label{prop:unifinj}
    The family of operators $\mathcal{P}_{\eps}$ is
    injective provided that $\eps \leq \eps_0$ is sufficiently small.
\end{prop}
    
\begin{proof}
Substitute $J_{abc} = \frac{1}{2} ( \eps^{e} _{\hspace{1ex} bc} F_{ae} + A_{b} g_{ac} - A_{c} g_{ab} )$ into the expression for $\mathcal{P}_{\eps}$ to obtain after some straightforward calculation
\begin{equation}
    \mathcal{P}_{\eps} (F, A) = \left(
    \begin{gathered}
	\eps^{e}_{\hspace{1ex} cb} \nabla^{c} F_{ae} + \eps^{e}_{\hspace{1ex} 
	ca} \nabla^{c} F_{be} - \nabla_{a} A_{b} - \nabla_{b} A_{a} + 
	\frac{2}{3} \nabla^{c} A_{c} g_{ab} \\[0.5ex]%
	\nabla^{b} F_{ab} + \eps^{bc}_{\hspace{1.5ex} a} \nabla_{b} A_{c}
    \end{gathered}
    \right)
    \label{eqn:fa}
\end{equation}
for $F$ and $A$.  Now suppose that $(F,A) \in H^{k-2, \beta-2} \big(
\symtf \times \oneform \big)$ and that $\mathcal{P}_{\eps}(F,A) = (0,0)$.  The key to showing $(F,A) = (0,0)$ is to find an inequality satisfied by $F$ and $A$ of the type needed to invoke Lemma \ref{lemma:contra}.   Begin with the identity $\eps^{e}_{\hspace{0.5ex} bc} F_{ae} + \eps^{e}_{\hspace{0.5ex} ab}
    F_{ce} + \eps^{e}_{\hspace{0.5ex} ca} F_{be} = 0$  that $F$ satisfies by virtue of being trace-free.  Then take its divergence with respect to $\nabla^{c}$ and substitute from
the second equation of $\mathcal{P}_{\eps}(F,A) = (0,0)$ to find
\begin{align}
    \eps^{e}_{\hspace{0.5ex} ca} \nabla^{c} F_{be} &= \eps^{e}_{
    \hspace{0.5ex} cb} \nabla^{c} F_{ae} + \eps^{e}_{ \hspace{0.5ex}
    ab} \eps^{st}_{\hspace{1ex} e} \nabla_{s} A_{t} \notag \\
    &= \eps^{e}_{ \hspace{0.5ex} cb} \nabla^{c} F_{ae} + \nabla_{a} 
    A_{b} - \nabla_{b} A_{a} \, .
    \label{eqn:calc2}
\end{align}
Next, substitute \eqref{eqn:calc2} into the first equation of 
$\mathcal{P}_{\eps}(F,A) = (0,0)$ and contracting with $\eps^{b}_{\hspace{0.5ex} uv}$ to get
 \begin{align}
     0 &= \eps^{b}_{\hspace{0.5ex} uv} \eps^{e}_{ \hspace{0.5ex} cb}
    \nabla^{c} F_{ae} - \eps^{b}_{\hspace{0.5ex} uv} \nabla_{b} A_{a}
    + \frac{1}{3} \nabla^{s} A_{s} \eps_{auv} \notag \\
    &= \nabla_{v} F_{au} - \nabla_{u} F_{av} -
    \eps^{b}_{\hspace{0.5ex} uv} \nabla_{b} A_{a} + \frac{1}{3}
    \nabla^{s} A_{s} \eps_{auv} \, .
    \label{eqn:calc3}
\end{align}
Now contract \eqref{eqn:calc3} with $\nabla^{v}$:
\begin{align}
    0 &= \Delta_{g} F_{au} - \nabla^{v} \nabla_{u} F_{av} -
    \eps^{b}_{\hspace{0.5ex} uv} \nabla^{v} \nabla_{b} A_{a} +
    \frac{1}{3} \nabla^{v} \nabla^{s} A_{s} \eps_{auv} \notag \\
    &= \Delta_{g} F_{au} - \nabla_{u} \nabla^{v} F_{av} - \frac{1}{2}
    \eps^{b}_{\hspace{0.5ex} uv} R^{v \hspace{1.5ex} s}_{
    \hspace{0.5ex} ba} A_{s} + \frac{1}{3} \nabla^{v} \nabla^{s} A_{s}
    \eps_{auv} - R^{s}_{u} F_{as} - R^{v\hspace{1.5ex}
    s}_{\hspace{0.5ex} ua} F_{sv} \notag \\
    &= \Delta_{g} F_{au} + \eps^{pq}_{\hspace{1.5ex} a} \nabla_{u}
    \nabla_{p} A_{q} + \frac{1}{3} \nabla^{v} \nabla^{s} A_{s}
    \eps_{auv} - \frac{1}{2} \eps^{b}_{\hspace{0.5ex} uv} R^{v
    \hspace{1.5ex} s}_{ \hspace{0.5ex} ba} A_{s} - R^{s}_{u} F_{as} -
    R^{v\hspace{1.5ex} s}_{\hspace{0.5ex} ua} F_{sv}
    \label{eqn:calc4}
\end{align}
after using the second equation of $\mathcal{P}_{\eps}(F,A) = (0,0)$ once again.  The advantage of this expression is that the only derivatives of $F$ appearing here are in the $\Delta F_{au}$ term.  A similar expression can be obtained for $A$ by computing the antisymmetric part of \eqref{eqn:calc4} and contracting with $\eps^{au}_{\hspace{1.5ex}t}$.  That is, 
\begin{align}
    0 &= \eps^{au}_{\hspace{1.5ex}t} \eps^{pq}_{\hspace{1.5ex} a}
    \nabla_{u} \nabla_{p} A_{q} + \frac{1}{3} \nabla^{v} \nabla^{s}
    A_{s} \eps_{auv} \eps^{au}_{\hspace{1.5ex}t} - \frac{1}{2}
    \eps^{au}_{\hspace{1.5ex}t} \eps^{b}_{\hspace{0.5ex} uv} R^{v
    \hspace{1.5ex} s}_{ \hspace{0.5ex} ba} A_{s} -
    \eps^{au}_{\hspace{1.5ex}t} R^{s}_{u} F_{as} -
    \eps^{au}_{\hspace{1.5ex}t} R^{v\hspace{1.5ex} s}_{\hspace{0.5ex}
    ua} F_{sv} \notag \\
    &= \Delta_{g} A_{t} - \nabla^{u} \nabla_{t} A_{u} + \frac{2}{3}
    \nabla_{t} \nabla^{b} A_{b} - R^{s}_{u} F_{as} \eps^{au}_{
    \hspace{1.5ex}t} - R^{s}_{t} A_{s} \notag \\
    &= \Delta_{g} A_{t} - \frac{1}{3} \nabla_{t} \nabla^{s} A_{s} -
    R^{s}_{u} F_{as} \eps^{au}_{ \hspace{1.5ex}t} \, .
    \label{eqn:calc5}
\end{align}

Since $F \in H^{k-2, \beta-2}$, the integral of $F^{au } \Delta_{g} F_{au}$ over $\R^{3}$ is well-defined and can be integrated by parts.  In what follows, $\mathcal{R}$ denotes a curvature quantity as in Lemma \ref{lemma:contra}.  Now,
\begin{align}
    0 &= \int_{\R^{3}} F^{au } \Delta_{g} F_{au} \dif \vol +
    \int_{\R^{3}} \eps^{pq}_{\hspace{1.5ex}a} \nabla_{u} \nabla_{p}
    A_{q} F^{au} \dif \vol + \int_{\R^{3}} \mathscr{R}(A,F) \dif \vol
    \notag \\
    &= \frac{1}{2} \int_{\R^{3}} \Delta_{g} \Vert F \Vert^{2} \dif
    \vol - \int_{\R^{3}} \Vert \nabla F \Vert^{2} \dif \vol \notag \\
    &\qquad + \int_{\R^{3}} \eps^{pq}_{\hspace{1.5ex}a} \nabla_{u}
    \nabla_{p} A_{q} F^{au} \dif \vol + \int_{\R^{3}} \mathscr{R}(A,F)
    \dif \vol \notag \\
    &= - \int_{\R^{3}} \Vert \nabla F \Vert^{2} \dif \vol +
    \int_{\R^{3}} \eps^{pq}_{\hspace{1.5ex}a} \nabla_{u} \nabla_{p}
    A_{q} F^{au} \dif \vol + \int_{\R^{3}} \mathscr{R}(A,F) \dif
    \vol\, ,
    \label{eqn:int1}
\end{align}
since $\int_{\R^{3}} \Delta_{g} u \dif \vol = 0$ for any $u \in
H^{k-2, \beta-2}(\R^{3})$.  But now,
\begin{align}
    \int_{\R^{3}} \eps^{pq}_{\hspace{1.5ex}a} \nabla_{u} \nabla_{p}
    A_{q} F^{au} \dif \vol &= - \int_{\R^{3}} 
    \eps^{pq}_{\hspace{1.5ex}a} \nabla_{p} A_{q} \nabla_{u} F^{au} 
    \dif \vol \notag \\
    &= \int_{\R^{3}} \eps^{pq}_{\hspace{1.5ex}a} \nabla_{p} A_{q}
    \eps^{rsa} \nabla_{r} A_{s} \dif \vol \notag \\
    &= \int_{\R^{3}} \Vert \nabla A \Vert^{2} \dif \vol -
    \int_{\R^{3}} \nabla^{s} A^{r} \nabla_{r} A_{s} \notag \\
    &= \int_{\R^{3}} \Vert \nabla A \Vert^{2} \dif \vol +
    \int_{\R^{3}} (\nabla^{s} A_{s} )^{2} \dif \vol + \int_{\R^{3}}
    \mathscr{R}(A) \dif \vol
    \label{eqn:int2}
\end{align}
using $\nabla_u F^{au} = - \eps^{rsa} \nabla_r A_s$ and repeated use of integration by parts.  Substituting \eqref{eqn:int2} into \eqref{eqn:int1} yields
\begin{equation}
    0 = - \int_{\R^{3}} \Vert \nabla F \Vert^{2} \dif \vol +
    \int_{\R^{3}} \Vert \nabla A \Vert^{2} \dif \vol - \int_{R^{3}} 
    (\nabla^{s} A_{s})^{2} \dif \vol + \int_{\R^{3}} 
    \mathscr{R}(A,F) \dif \vol \, .
    \label{eqn:int3}
\end{equation}

Similar calculations involving the equation \eqref{eqn:calc5} for
$\Delta_{g} A_{t}$ yield the identity
\begin{equation}
    0 = - \int_{\R^{3}} \Vert \nabla A \Vert^{2} \dif \vol + 
    \frac{1}{3} \int_{\R^{3}} (\nabla^{s} A_{s} )^{2} \dif \vol + 
    \int_{\R^{3}} \mathscr{R}(A) \dif \vol \, .
    \label{eqn:int4}
\end{equation}
Adding one third of equation \eqref{eqn:int3} to equation \eqref{eqn:int4} allows the divergence term to cancel and leads to the identity
\begin{equation}
    0 = - \int_{\R^{3}} \left( \frac{1}{3} \Vert \nabla F \Vert^{2} + 
    \frac{2}{3} \Vert \nabla A \Vert^{2} \right) \dif \vol + 
    \int_{\R^{3}} \mathscr{R}(A,F) \dif \vol \, .
    \label{eqn:int5}
\end{equation}
Now $\mathscr{R}(A,F) \leq C \Vert \mathscr{R} \Vert ( \Vert F \Vert^{2} + \Vert A \Vert^{2})$ where $\Vert \mathscr{R} \Vert$ denotes the supremum norm of curvature coefficients and $C$ is a general numerical constant.  Consequently, one obtains the estimate
\begin{equation}
    0 = - \int_{\R^{3}} \left( \Vert \nabla F \Vert^{2} + \Vert \nabla
    A \Vert^{2} \right) \dif \vol + C \int_{\R^{3}} \Vert \mathscr{R}
    \Vert \left( \Vert A \Vert^{2} + \Vert F \Vert^{2} \right) \dif
    \vol \, .
    \label{eqn:int6}
\end{equation}
Lemma \ref{lemma:contra} along with the simple estimate $\big\Vert \nabla \Vert T \Vert \, \Vert^{2} \leq C \Vert \nabla T \Vert^{2}$ (by Cauchy-Schwarz) can now be applied to conclude that $(F,A) = (0,0)$ when $\Vert h \Vert_{H^{k,\beta}}$ is sufficiently small.  But since $h$ depends smoothly on $(T, \bar{T}, \phi)$, and vanishes when these quantities are zero, the operator $\mathcal{P}_{\eps}$ is injective when $\eps$ is sufficiently small. 
\end{proof}

\begin{prop}
	For every $0 \leq \eps \leq \eps_0$, the operator $\mathcal{P}_{\eps}$ is elliptic.
\end{prop}

\begin{proof}

The symbol of $\mathcal{P}_\eps$, rewritten in terms of the isomorphism of Lemma \ref{lemma:jacdecomp} is given by
$$(F,A) \mapsto \left(
    \begin{gathered}
	\eps^{e}_{\hspace{1ex} cb} \xi^{c} F_{ae} + \eps^{e}_{\hspace{1ex} 
	ca} \xi^{c} F_{be} - \xi_{a} A_{b} - \xi_{b} A_{a} + 
	\frac{2}{3} \xi^{c} A_{c} g_{ab} \\[0.5ex]%
	\xi^{b} F_{ab} + \eps^{bc}_{\hspace{1.5ex} a} \xi_{b} A_{c}
    \end{gathered}
    \right) 
$$
for any non-zero $\xi \in \R^3$.  If one follows the same algebraic steps as were used in the previous proposition, one ends up with the two equations
\begin{align*}
    0 &= \Vert \xi \Vert^2 F_{au} - \xi^{v} \xi_{u} F_{av} -
    \eps^{b}_{\hspace{0.5ex} uv} \xi^{v} \xi_{b} A_{a} +
    \frac{1}{3} \xi^{v} \xi^{s} A_{s} \eps_{auv} \\
    0 &=  \Vert \xi \Vert^2 - \frac{1}{3} \xi_t \xi^s A_s \, .
\end{align*}
(These are the analogs of equations \eqref{eqn:calc4} and \eqref{eqn:calc5}, where curvature terms do not appear since commuting derivatives reduces to commuting multiplication by $\xi$ at the level of the symbol.)  Contracting the second equation against $\xi^t$ yields $\frac{2}{3} \Vert \xi \Vert^2 \xi^t A_t = 0$ or $\xi^t A_t = 0$.  If this is substituted back into the first equation, one has $\Vert \xi \Vert^2 A_t = 0$, or simply $A_t = 0$.  In a similar manner, one then concludes that $F_{au} = 0$.

The symbol of $\mathcal{P}_{\eps}$ is thus injective.  But since it is a map from one 8-dimensional space of tensors to another, it must be surjective as well.  Consequently, $\mathcal{P}_\eps$ is elliptic.
\end{proof}

\begin{cor}
	The family of operators $\mathcal{P}_\eps$ is uniformly injective, provided $\eps$ is sufficiently small.
\end{cor}

\begin{proof}
The injectivity of $\mathcal{P}_\eps$ together with the elliptic estimate yields an estimate of the form $\Vert u \Vert \leq C \Vert \mathcal{P}_\eps u \Vert$ in the appropriate norms.  Since the coefficients of $\mathcal{P}_\eps$ depend smoothly on $\eps$ and $\mathcal{P}_0$ is not degenerate, the constant $C$ can be made independent of $\eps$.  But this is the uniform injectivity of the operators $\mathcal{P}_\eps$.
\end{proof}

\paragraph*{Transversality and Completion of the Proof.}  The equation for $J$ in \eqref{eqn:j} is schematically of the form $\mathcal{P}_{\eps}(J) = \Lambda_{\eps}$, where $\Lambda_{\eps}$ depends on $\eps$ is the error term.  However, for each $\eps \leq \eps_{0}$, $\Lambda_{\eps}$
belongs to the fixed, finite dimensional subspace 
$$\mathcal{C}' = \left\{\eta \mu_{ij} \dif x^{i} \otimes \dif x^{j} \,
: \, \mu_{ij} \in \R \right\} \times \left\{ \eta
\bar{\lambda}_{i} \dif x^{i} \, : \, \bar{\lambda}_i \in \R
\right\} \, ,$$
of $H^{k-3, \beta-3} \big( \symtf \times \oneform \big)$.  But the following transversality result makes this impossible unless $\Lambda_\eps = 0$.

\begin{prop} 
	The family of operators $\mathcal{P}_{\eps}$ satisfies $\mathcal{C}' \cap \mathrm{Im}(\mathcal{P}_{\eps}) = \{ (0,0) \}$ provided $\eps$ is sufficiently small.
\end{prop}

\begin{proof}

Apply the divergence theorem to the equation $\mathcal{P}_{0} (J) = \eta (\mu, \bar{\lambda})$.  Integrating over $\R^{3}$ gives zero on the left hand side, while by construction, integrating the right hand side yields $(\mu_{ab}, \bar{\lambda}_{a})$. Thus the conclusion of the theorem holds for $\mathcal{P}_0$.  Consequently, by Proposition \ref{prop:stab} and the uniform injectivity of $\mathcal{P}_{\eps}$, the image of $\mathcal{P}_{\eps}$ remains transverse to $\mathcal{C}'$ for sufficiently small $\eps$. 
\end{proof}

The equation \eqref{eqn:j} states that $\eta (\mu, \bar{\lambda}) \in \mathcal{C}' \cap \mathrm{Im} (\mathcal{P}_\eps)$.  Thus $(\lambda, \mu)$ must vanish, and $J$ must satisfy $\mathcal{P}_\eps(J) = (0,0)$.  Thus one can invoke the uniform injectivity of $\mathcal{P}_{\eps}$ once again to conclude that $J$ itself must vanish.  The consequence of vanishing $J$, $\bar{\lambda}$ and $\mu$ is that the equation for $\Gamma$ to reads $\frac{1}{2} \big( \Delta_{g} \Gamma_{b} + R^{a}_{b} \Gamma_{a} \big) = - \eta \lambda_{b}$.  The vanishing of both $\Gamma$ and $\lambda$ can now be shown in the same way as above.  That is, since  $-\eta \lambda$ belongs to the fixed, finite-dimensional subspace $\mathcal{C}'' \equiv \{ \eta \lambda_{i} \dif x^{i} \, : \, \lambda_{i} \in \R \}$ and $\Delta_g - \Ric$ is uniformly injective and elliptic, all that is needed is the transversality of $\Delta_\delta$ to $\mathcal{C}''$ by Proposition \ref{prop:stab}.  But this is again a simple consequence of the divergence theorem.  Consequently, $\lambda = 0$ and so $\Gamma = 0$ by uniform injectivity.

This completes the proof of Theorem B. In combination with Theorem A, this yields a construction of perturbative solutions of the extended constraint equations and concludes the proof of the Main Theorem.

\hfill \QED

\newpage

\renewcommand{\baselinestretch}{1}
\normalsize

\bibliography{grav}
\bibliographystyle{amsplain}

\end{document}